\documentclass{nature}
\usepackage{amsmath,amssymb, amstext}
\usepackage{amsmath}
\usepackage{graphicx}
\usepackage{bbold}
\usepackage{amssymb}
\usepackage{dcolumn}
\usepackage{color}
\usepackage{bm}
\usepackage{pdfpages}
\usepackage[normalem]{ulem}

\newcommand{\beq}{\begin{equation}}
\newcommand{\eeq}{\end{equation}}
\newcommand{\bOne}{{\boldsymbol 1}}

\usepackage{soul}
\usepackage{url}
\usepackage[colorlinks = true,
linkcolor = blue,
urlcolor  = red,
citecolor = blue,
anchorcolor = red]{hyperref}

\begin{document}
\title{Signatures of emergent surface states across a displacive topological phase transition in Bi$_4$I$_4$}

\author{Deep Singha Roy$^{1,\S}$, Sk Kalimuddin$^{1,\S}$, Subrata Pachhal$^{2,\$}$, Saikat Mondal$^{2,\$}$, Soham Das$^{1}$, Sukanya Jana$^{1}$, Arnab Bera$^{1}$,  Satyabrata Bera$^{1}$, Tuhin Debnath$^{1}$, Ankan Bag$^{1}$, Souvik Pramanik$^{1}$, Sudipta Chatterjee$^{1}$, Sanjib Naskar$^{1}$, Shishir Kumar Pandey$^{3,4}$\footnote{e-mail: shishir.kr.pandey@gmail.com }, Adhip Agarwala$^{2}$\footnote{e-mail: adhip@iitk.ac.in }, Mintu Mondal$^{1}$
\footnote{e-mail: sspmm4@iacs.res.in}}

\maketitle
\begingroup
\renewcommand\thefootnote{}\footnote{$^{\S}$ These authors contributed equally to this work}%
\addtocounter{footnote}{-1}%
\renewcommand\thefootnote{}\footnote{$^{\$}$ These authors contributed equally to this work}%
\addtocounter{footnote}{-1}%
\endgroup

\begin{affiliations}
\item School of Physical Sciences, Indian Association for the Cultivation of Science, Jadavpur, Kolkata 700032, India
\item Indian Institute of Technology Kanpur, Kalyanpur, Uttar Pradesh 208016, India
\item Department of General Sciences, Birla Institute of Technology and Science, Pilani-Dubai Campus, Dubai International Academic City, Dubai 345055, UAE
\item Department of Physics, Birla Institute of Technology and Science, Pilani, Hyderabad Campus, Jawahar Nagar, Kapra Mandal, Medchal District, Telangana 500078, India
\end{affiliations}

\begin{center}
    \textbf{Abstract}
\end{center}
\noindent\textbf{Topological phase transitions involving crystalline symmetry breaking provide a fertile ground to explore the interplay between symmetry, topology, and emergent quantum phenomena. Recently discovered quasi-one-dimensional topological material, Bi$_4$I$_4$, has been predicted to host topologically non-trivial gapless surfaces at high temperature, which undergo a finite temperature phase transition to a low temperature gapped phase. Here we present experimental signatures of this room temperature phase transition from a high-temperature $\beta$-phase with a surface state to a gapped $\alpha$-phase  hosting hinge states. Using real-space current mapping and resistance fluctuation spectroscopy, we identify signatures of a displacive topological phase transition mediated by a first-order thermodynamic structural change. Near the emergence of $\beta$-phase, we observe pronounced telegraphic noise, indicating fluctuating phase domains with topological surface states. The spatially resolved current map reveals electron transport via the gapless surface states in the $\beta$-phase, which vanishes upon transitioning to the $\alpha$-phase with localized conduction channels (or hinge modes). Our experimental results, supported by first principles estimates and effective theory of a topological displacive phase transition, establish Bi$_4$I$_4$ as a candidate material showing intricate interplay of classical thermodynamic phase transitions with topological quantum phenomena.}

\addtocontents{toc}{\protect\setcounter{tocdepth}{-10}}
\section{Introduction}
Three-dimensional topological insulators (3D TIs) are characterized by a non-trivial bulk and topologically protected conducting surface states\cite{Fu2007PRL,Qi2008PRB,Hsieh2008,Chen2009Science}. These surface states arise due to the nontrivial topological invariants of the bulk state and are protected by the time-reversal symmetry\cite{hasan2010colloquium, ShouChengRevModPhy, Bansil_RevModPhys}. The robustness of these gapless states makes 3D TIs promising materials for applications in low-dissipation electronics and topologically protected quantum information processing\cite{MajoranazeromodesPRB,HOTI_Bismuth,TopologicalSuperconductivity}. Another class of interesting topological materials in $d$ spatial dimensions contains boundary modes which reside in $(d-2)$ dimensions, notably called higher-order topological insulators (HOTIs)\cite{BBH_hoti_2017, HOTI_sciAdvschindler2018}. These contain one-dimensional conducting states, often at the hinges, even while the bulk and the surfaces remain insulating. Albeit extensive theoretical predictions of realizing HOTIs in a host of compounds, viz.~SnTe, EuIn$_2$As$_2$ \cite{HOTI_sciAdvschindler2018, NguyenPhysRevB.105.075310, Pyrochlore_PhysRevLett, Symmetry-enforced_NatPhys, HOTI_EuIn2As2}, concrete experimental realizations have been far and few \cite{HOTI_Bismuth,PRX_3D_QSHI_Bi4I4yu2024,Bi4Br4_HOTISTM_Hasan2022evidence}. Direct transport signatures of such edge modes have been challenging due to lattice imperfections and their susceptibility to localization. Such localization of surface modes also occurs in weak TIs, where even though the boundary modes exist in selective $(d-1)$ surfaces, their topological protection is often limited by the existence of pristine crystalline symmetries. It is in this context where the recent prediction and spectroscopic signatures of topological edge modes in quasi-one-dimensional Bi$_4$X$_4$ (X = I, Br) have been of immense interest \cite{Bi4Br4_HOTISTM_Hasan2022evidence, Bi4Br4_PNAS2019, Bi4Br4_Rotaion_protectd_hsu2019purely,  zhong2025coalescence,Quantum_Trmsport_Bi4Br4_hossain2024quantum, PRX_3D_QSHI_Bi4I4yu2024,PhysRevX.11.031042,2024_mass_acquisition_PRL,noguchi2019_WTI,noguchi2024robust}. Their layered crystal structure enables mechanical exfoliation down to thin flakes. Their building blocks, consisting of Bi chains aligned along a single crystallographic axis, impart strong structural anisotropy and an inherently one-dimensional character to the material. Notably, their topological properties are sensitive to interlayer coupling and stacking order of Bi$_4$X$_4$ layers, enabling distinct topological phases across the polymorphs. Recent scanning tunneling spectroscopy (STS) and ARPES experiments have reported edge-localized modes in Bi$_4$Br$_4$, providing tantalizing evidence of topological character in this material class\cite{Bi4Br4_HOTISTM_Hasan2022evidence,Bi4Br4_Rotaion_protectd_hsu2019purely,zhong2025coalescence}.  This places these quasi-one-dimensional materials at a unique juncture where they are predicted to reflect subtle topological character in their different crystalline structures. 

Among these quasi-one dimensional compounds, Bi$_4$I$_4$, undergoes a unique lattice transition at ambient temperatures where even though the space group symmetry remains unchanged but the unit cell doubles in the low temperature $\alpha$ phase\cite{PhysRevX.11.031042,yoon_arxiv_2020,npjQuantum_topologicalphasetransition_Bi4i4,2024_mass_acquisition_PRL}. This structural transition drastically modifies the band structure of the itinerant electrons \cite{autes2016novel_WTI, PRX_3D_QSHI_Bi4I4yu2024, PhysRevX.11.031042}. While the $\beta$-phase is understood as a stacking of quantum spin-Hall systems in the ${\bf c}$ direction, rendering the (100) surface with a gapless Dirac state\cite{npjQuantum_topologicalphasetransition_Bi4i4,PRX_3D_QSHI_Bi4I4yu2024,PhysRevX.11.031042}, the low-temperature $\alpha$ phase is known to open a gap of $\sim 35$ meV\cite{PhysRevX.11.031042}.  Extensive investigations into the surfaces of both $\alpha$ and $\beta$ phases have been reported either through imaging and spectroscopic signatures\cite{noguchi2019_WTI, PRX_3D_QSHI_Bi4I4yu2024}; direct transport signatures of such non-trivial surfaces and edge states remain lacking\cite{Bi4I4_GateTunable_liu2022}. In this work, we address this lacuna by mapping of the surface currents in the $\beta$ phase and discerning the unique signatures of the boundary modes of $\alpha$ phase in the noise spectroscopy of the system. Further using both first principles and an effective tight-binding model of itinerant electrons interplaying with a phonon-mediated displacive lattice transition, we show the finite temperature phase transition of Bi$_4$I$_4$ is an unique example of a displacive topological phase transition.

\begin{figure*}[ht!]
\centerline{\includegraphics[width=1.0\textwidth]{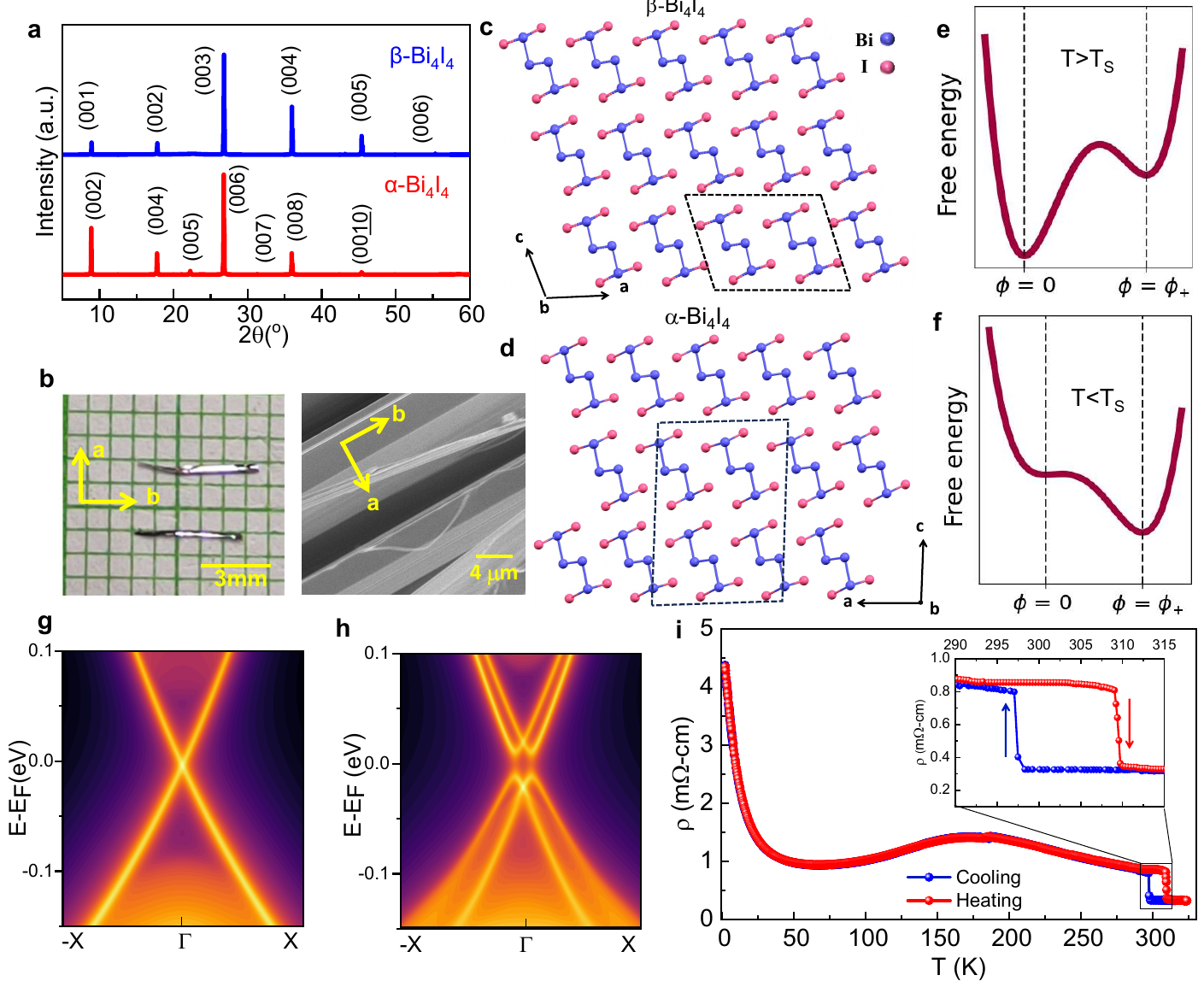}}
\caption{\textbf{Room temperature topological phase transition.} \textbf{a,} Room-temperature XRD data for Cu K$\alpha$ X-ray source projected on {\bf ab} plane of Bi$_4$I$_4$ crystal for both $\beta$ (top) and $\alpha$ (bottom) phase. \textbf{b,} Image of as-grown single crystals on a millimeter-scale graph paper (left) and its scanning electron microscopy image (right). \textbf{c-d,} Schematic representation of Bi$_4$I$_4$ crystal structure inferred from Laue-diffraction for $\beta$ and $\alpha$ phases respectively. The projected unit cell as demarcated by the dashed box shows doubling of unit cell along \textbf{c} axis in $\alpha$ phase. \textbf{e-f,} Schematically shown free energy landscape of $\beta$ and $\alpha$ phase with their local minima at $\phi=0$ and $\phi= \phi_{+}$ , respectively, following $\phi^3$-theory of displacive first-order phase transition. \textbf{g}, First principle calculation shows Dirac-like gapless surface state at (1 0 0) surface of $\beta$-Bi$_4$I$_4$. \textbf{h}, Gapped nature at $\Gamma$ point near the Fermi level of ($\Bar{2}$ 0 1) surface in the $\alpha$-phase. \textbf{i,} Temperature-dependent DC resistivity data for cooling (blue) and heating (red) cycles demonstrating a pronounced first-order phase transition near room temperature with a hysteresis. The inset shows the thermal hysteresis between cooling and heating cycles.}
\label{Figure1}
\end{figure*} 

\section{The displacive topological transition}
Single crystals of Bi$_4$I$_4$ are grown using the chemical vapor transport (CVT) technique. Figure \ref{Figure1}a shows the XRD data with (0 0 $l$) orientation peaks for $\beta$ (blue) and (0 0 $2l$), along with weak (0 0 5), (0 0 7) peaks in the $\alpha$ (red) phases of Bi$_4$I$_4$ respectively. This confirms the crystallinity and lowering of symmetry from the $\beta\rightarrow\alpha$ phase. Bi$_4$I$_4$ crystallizes in a monoclinic lattice structure with space group C2/m. Figure \ref{Figure1}b depicts the image of as-grown Bi$_4$I$_4$ crystals and their scanning electron microscopy image. Both $\alpha$ and $\beta$ phases consist of quasi 1D chains of Bi$_4$I$_4$ units stretched along the {\bf{b}} axis with Bi-Bi metallic bonding\cite{Bi4I4_GateTunable_liu2022, PhysRevX.11.031042}. The unit cell of Bi$_4$I$_4$ projected onto the {\bf{a-c}} plane is schematically represented in Figure~\ref{Figure1}c,d. Despite having the same space group, the $\alpha$ (low-temperature) phase contains doubly stacked chains, whereas $\beta$ (high-temperature) phase comprises single stacking of Bi$_4$I$_4$ units. Thus in the $\alpha$ phase, {\bf{c}} gets doubled  ($10.5\text{Å}\rightarrow19.97~\text{Å}$) and so does the corresponding cell volume ($636~\text{Å}^{3}\rightarrow 1257~\text{Å}^{3}$). This alters the stacking order and each Bi$_4$I$_4$ monolayer shifts along {\bf{a}}-direction. Therefore, the softening of the phonon mode leading to this lattice transition allows one to describe it using the $\phi^3$-theory of displacive first-order phase transition\cite{Krumhansi_PRB_1989}. In the high temperature phase while Figure~\ref{Figure1}e shows the free energy landscape for a scalar field $\phi$, representing the phononic order parameter has a minima at $\phi=0$, at a temperature below a critical temperature $T_S$ the field value suddenly changes to a finite positive value $\phi_{+}$ signifying the first-order transition (see Figure~\ref{Figure1}f). 

In Figure \ref{Figure1}g,h, we present the plots of surface states calculated using the Green's function approach, as implemented in WannierTools package~\cite{wt}, for $\beta$ and $\alpha$-phases of Bi$_4$I$_4$, respectively. Consistent with the experimental finding of conducting surface states, our (1 0 0) surface states of the $\beta$ phase calculated along -$X \leftarrow \Gamma \rightarrow X$ also show conducting behavior with Dirac-like linear crossings at the Fermi level, as shown in Figure~\ref{Figure1}g. Given the difference between the $\beta$-angle of $\beta$ and $\alpha$ phases, the (1 0 0) surface becomes the ($\Bar{2}$ 0 1) surface in the $\alpha$ phase. In Figure~\ref{Figure1}h, we plot this ($\Bar{2}$ 0 1) surface states in which the gapped nature of the bands at the crossing point ($\Gamma$) is evident, which again is consistent with the experimental finding. 
It is interesting to note that the gap of the surface state in the $\alpha$ phase is $\sim 25~$meV, while the bulk band gaps are $\sim$130~meV ($\alpha$-phase) and $\sim$60 meV ($\beta$-phase). This implies that any transport signatures at ambient temperatures are expected to result from only the surface electrons.

This unique topological phase transition from $\beta$ to $\alpha$ and vice-versa, occurs in the vicinity of room temperature ($T_S\sim300$ K), is evidenced in the resistivity with a hysteretic window $\sim10$ K, as shown in Figure~\ref{Figure1}i. Apart from this room temperature phase transition, resistivity exhibits a broad hump at around $150$ K followed by a sharp upturn below $100$ K, indicating an insulator-like behavior Figure~\ref{Figure1}i. This broad hump is also anticipated to be a consequence of temperature-induced Lifshitz transition in Bi$_4$X$_4$\cite{Bi4Br4Lifshitz_acsNano, Bi4IBr4Lifshitz}. The question arises whether the gapless surface states of the $\beta$ phase, which are predicted to be of topological origin, directly contribute to electron transport. 

\section{Real space nanoscale map of local currents - signatures of transport via surface states}
We therefore investigate the local conductance of two topologically nontrivial phases of Bi$_4$I$_4$ \cite{PhysRevX.11.031042,PRX_3D_QSHI_Bi4I4yu2024} using the conductive Atomic Force Microscopy (c-AFM) technique. The experimental scheme for c-AFM measurement is schematically represented in Figure S6 of the Supplementary Information. Given $\beta$-Bi$_4$I$_4$ phase is a weak TI\cite{autes2016novel_WTI,noguchi2019_WTI}, it is known to host conducting states on the side surfaces like (100),($\Bar{1}$00)\cite{WTI_Composite_Wyel_zhang2016,npjQuantum_topologicalphasetransition_Bi4i4,yoon_arxiv_2020}. This allows us to distinguish the two topological phases across the transition via the spatial imaging of local current density. We exfoliate a pristine single crystal in a glove box in an Argon environment and transfer flakes onto a p-type Si wafer. Despite having an apparent quasi-1D character, Bi$_4$I$_4$ crystals typically cleave along the (001) direction (i.e., {\bf a-b} plane)\cite{WTI_Composite_Wyel_zhang2016}. A higher exfoliation energy along the \textbf{a}-direction compared to the \textbf{c}-direction favors this exfoliation\cite{WTI_Composite_Wyel_zhang2016}. 

\begin{figure*}[ht!]
\centerline{\includegraphics[width=1.0\textwidth]{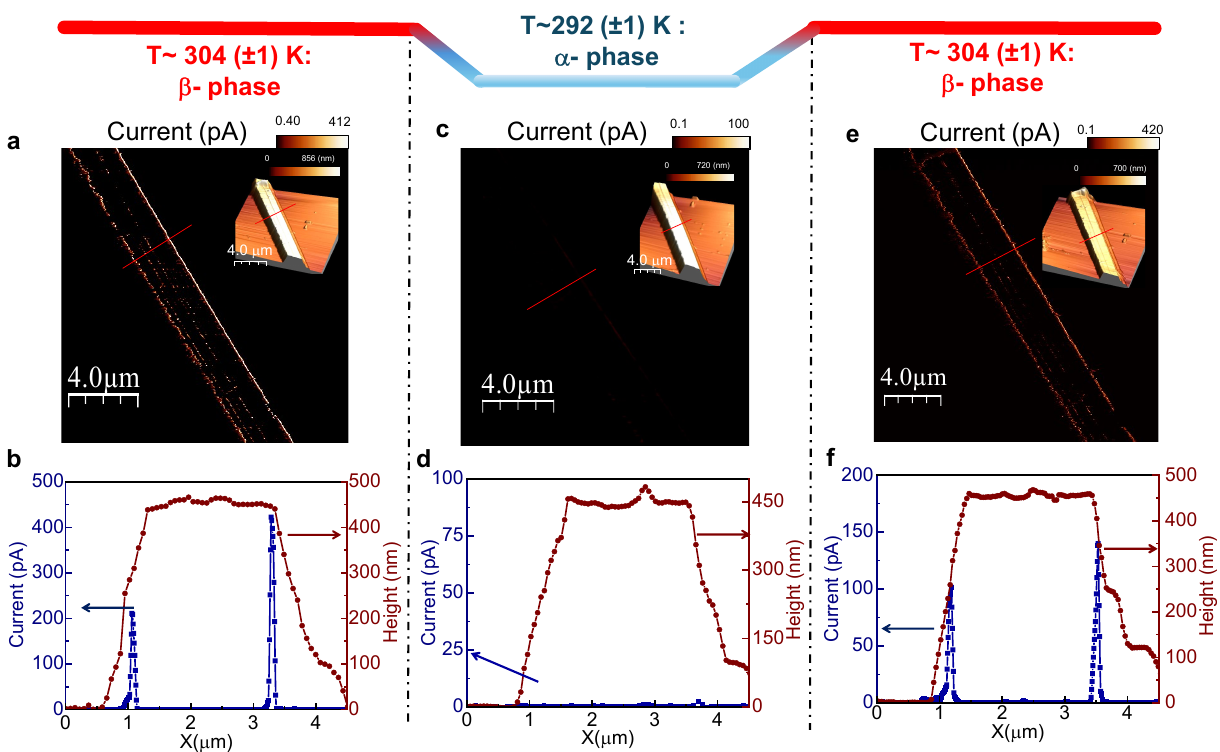}}
\caption{\label{mf3.6}\textbf{Evidence of vanishing surface current across the transition of Bi$_4$I$_4$ flakes.} \textbf{a-b,} Spatial current map (topography in the inset) of a exfoliated Bi$_4$I$_4$ flake (thickness $\sim~400~$nm) $\beta$-phase at $T\sim304$ K and plot of current variation (left) and flake thickness (right ) along the respective lines, respectively. The scans are recorded with a bias voltage of $1$ V DC and a standard scan rate of $1$ Hz. \textbf{c-d,}  Spatial current map (topography in the inset) of the same flake in $\alpha$-phase at $T\sim292$ K ($< T_S$) showing no edge current. \textbf{e-f,}  Current and topography maps after reheating the flake to the $\beta$-phase at $T \sim 304$ K (heated $> T_S$ and stabilized around $304$ K). The reappearance of edge current confirms its reversible and robust nature.}
\label{Fig:CAFM}
\end{figure*} 

The topographic image of the flake is shown in the inset of Figure \ref{Fig:CAFM}a. We next measure the current distribution of the flake in the $\beta-$phase as shown in Figure \ref{Fig:CAFM}a. Interestingly, the current map reveals significantly large conduction along the step edges located at the side surfaces, consistent with the predictions that $\beta-$Bi$_4$I$_4$ hosts gapless states in four of its geometric planes other than (001), (00$\overline{1}$). The line scan of height profile and the corresponding current on a specific cross-section (shown by a red dashed line in Figure \ref{Fig:CAFM}a) is shown in Figure \ref{Fig:CAFM}b. Our First-principles calculations also support this observation that the $\beta$-phase has gapless states at all its surfaces except the top and bottom surfaces parallel to \textbf{ab}(001) plane (see Fig. S5). Now, as we reduce the sample temperature to $\approx$~292~K, such that the material transits to the $\alpha $-phase which hosts no such surface states. Consistent with this expectation, the c-AFM measurements effectively capture a suppression of the currents at (100) surface of the sample as shown in Figure~\ref{Fig:CAFM}c,d. Interestingly, in a thermal cycle where the sample is re-heated to the $\beta$ phase, the (100) surface again starts conducting, as seen in Figure~\ref{Fig:CAFM}e,f. We have also observed signatures of all the side surfaces like (100),(010) conduct in another flake, which is presented in the supplementary information(See Figure S6). We have checked this hysteretic cycle, and the same behaviour in the current distribution is observed (See Figure S7). The above observations thus show the change of the topological character as well as charge transport signature of Bi$_4$I$_4$ across the first-order lattice transition. 

\section{Two-level fluctuations in resistance noise}

Low-frequency 1/f noise spectroscopy has emerged as a promising experimental technique for investigating electronic transport mechanisms and phase transitions in solid-state systems\cite{weissmanRMP1981,kogan_noise}. Notably, the temporal resistance fluctuations are sensitive to underlying changes in the scattering mechanism associated with the low-energy dynamics\cite{Koushik_R.2013,Kalimuddin_PRL}. Moreover, this technique can also differentiate between transport channels in materials, for example, localized modulation of edge/surface conduction (or filamentary conduction) and diffusive transport, by capturing their unique signatures, including random telegraphic signals\cite{1tTaS2Hidder2021,raquet2000noise}. Here, we use this technique for Bi$_4$I$_4$ to understand the subtle changes in the transport mechanism across the displacive resistive transition.

\begin{figure*}[ht!]
\centerline{\includegraphics[width=1.02\textwidth]{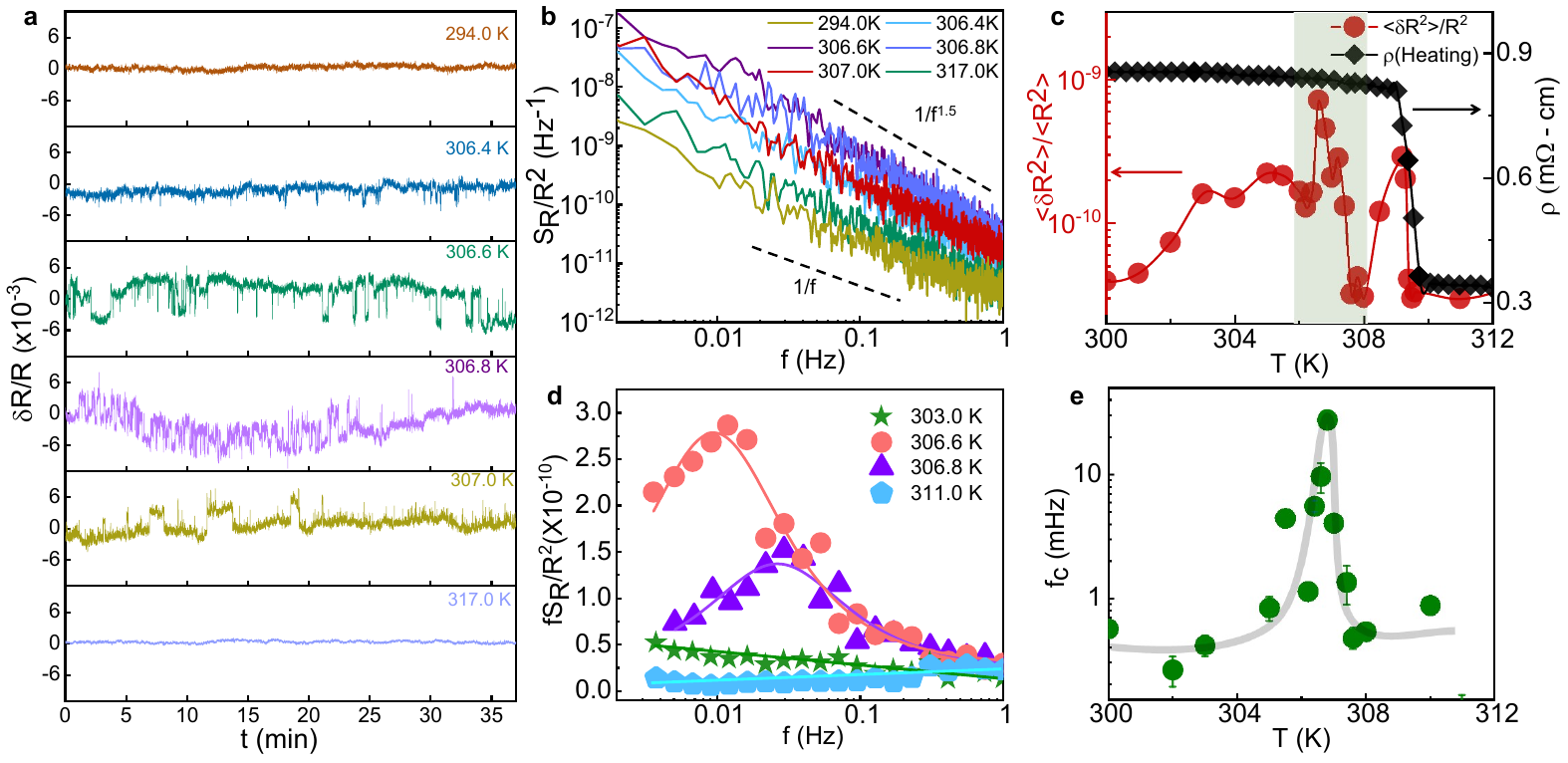}}
\caption{\label{Figure 4}\textbf{Two level fluctuations near topological phase crossover.} \textbf{a,} Time series of normalized resistance fluctuations at a few representative temperatures in the very vicinity of the phase transition in heating cycle. \textbf{b,} Power spectral density of the representative time series data. A clear deviation from the 1/$f$ behavior is observed near the TLF region. \textbf{c,} Enhancement in integrated normalized noise spectra (noise variance) occurs at around 306.6~K. \textbf{d,} Plot of $f$S$_R$/R$^2$ at selected temperatures. \textbf{e,} Extracted corner frequency of the Lorentzian component (f$_c$) shows a sharp enhancement at around 306.8~K.}
\label{Fig:Noise1}
\end{figure*} 

Figure \ref{Fig:Noise1}a shows the time series plot of the detrended resistance fluctuations at a few representative temperatures that include those very close to $T_S$. Also, a zoomed-in representation of the timeseries data is shown in the supporting information (Figure~S9). The power spectral density (PSD) of the digitally filtered resistance fluctuations within a bandwidth of $1$ mHz - $1$ Hz is numerically calculated employing Welch's Periodogram method \cite{Arindam2004Arxiv,Kalimuddin_PRL}. The normalized PSD data exhibits typical $1/f^{\alpha}$ behavior with an inherent Lorentzian for temperature values close to $T_S$ where $\alpha$ represents the noise exponent (see Figure \ref{Fig:Noise1}b). Interestingly, the Lorentzian feature in the PSD is consistent with the two-level fluctuations (TLF) in the timeseries data as well. This random telegraphic noise (or TLF) emerges just prior to the  $T_S$ ($\sim309$ K), where resistance shows a jump in its absolute value. This is rather surprising given our system is a three-dimensional bulk material; while telegraphic noise has earlier been observed in thin films \cite{RTN_MOSFET2018,RTN_PCMOAKRbid2003low,raquet2000noise}, devices\cite{1tTaS2Hidder2021, NbSe2_Bid_kundu2017quantum}, semiconductors, and their heterostructures ~etc.~often due to trap states \cite{Semiconductor_Two-parameter_Singnalmachlup1954noise,Trap_Induced_MobilityFlucgauthier2023universality,RTN_MOSFET2018, raquet2000noise,RTN_In_Semiconductor_Devicestataridou2022model, Benchmarking.noise }. More interestingly, we find such a signal only close to the phase transition $T_S$, reflecting that the origin of the telegraphic signal is intricately related to the nature of the phases at the first-order transition. This also hints at the possibility of transport dominated through side surfaces in the $\beta$-phase. 

We further estimate the relative noise variance, an integrated noise PSD spectrum within a chosen band, given by $\frac{\langle\delta{R^{2}} \rangle}{\langle R^{2}\rangle} =\int_{f_{min}}^{f_{max}} \frac{S_{R}(f)}{R^{2}}df$\cite{Koushik_R.2013,Benchmarking.noise}. Clearly, relative variance as shown in Figure \ref{Fig:Noise1}c (left) exhibits a sharp rise near $T_S$ as well as near the demarcated region where TLF is witnessed. We further rescale the PSD i.e., $f \frac{S_R}{R^2}$ to extract $f_C$ and plot the same (See Figure \ref{Fig:Noise1}d,e) where, $S_{R}/R^{2}=A/f+Bf_{c}/(f^{2}+f_{c}^{2})$. Notably, $f_C$ shows a pronounced peak near 306.6~K. Given that the origin of telegraphic noise is often attributable to effective two-level systems, we next estimate the typical time scales for such two-level fluctuations. We note these two level switchings are slow (orders of seconds), which suggests these levels are separated by small energy scales. Now, whether low-frequency noise possesses the signatures of topological surface states or not. As our effective theoretical description demonstrates that the charge transport becomes very slow as they have to tunnel through the large domains of $\alpha$-phase (say $\sim$ 400 nm) from one $\beta$ gapless conducting state to another. 

\section{Theoretical framework}
\begin{figure*}
    \centering \includegraphics[width=1\linewidth]{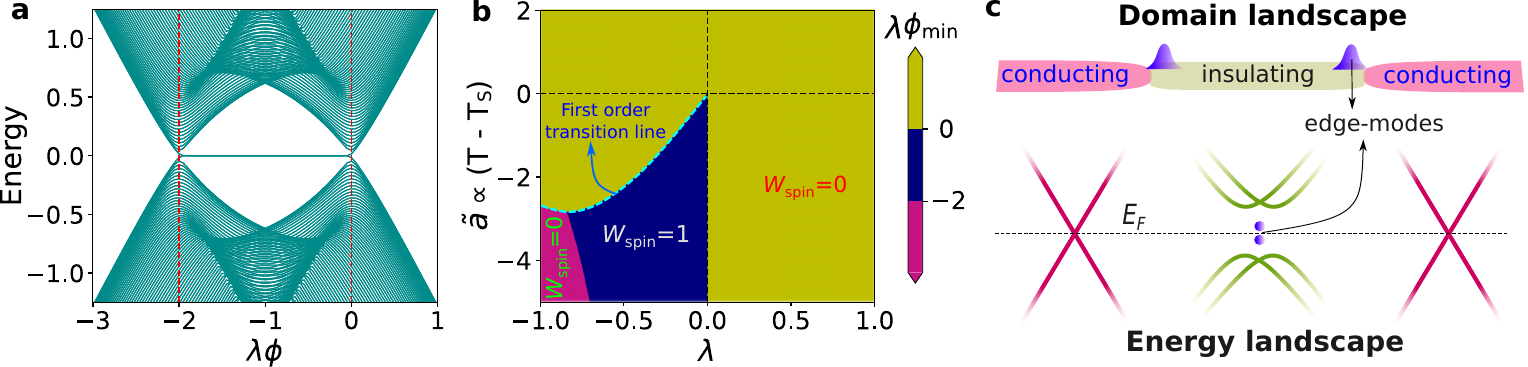}
    \caption{{\textbf{Edge-modes as fluctuation channel}}: \textbf{a,} Energy spectrum of the effective one-dimensional Hamiltonian in the $\alpha$-phase host zero-energy modes within the bulk gap for the parameter regime $-2<\lambda \phi<0$. \textbf{b,} The global minima $\phi_{\text{min}}$ of the total free energy density (with both electronic and phononic part) multiplied with $\lambda$ is shown in the $\lambda$ - $\tilde{a}$ parameter space.  The region where $\lambda \phi_{\text{min}}$ is within the range $(-2, 0)$, electronic Hamiltonian become topological. \textbf{c,} Effective one-dimensional domain and the corresponding energy landscape for the resistivity fluctuation near the transition.}
    \label{fig_flucchannel}
\end{figure*}

Having established the displacive first-order transition and the concomitant transition of the topological electronic structure, we now devise a minimal theoretical model to capture this phenomenology. While a scalar field $\phi$ models the phononic modes that lead to a structural transition, their coupling to itinerant fermions can modify the effective band structure. The effective Lagrangian for the $\phi$ field is given by 
\begin{equation}
    {\mathcal {L}} = \frac{a}{2} {\phi^2}- \frac{1}{3} \phi^3 + \frac{1}{4} \phi^4
\end{equation}
where $a=\tilde{a} + 2/9$ and $\tilde{a} \propto (T - T_S)$, such that for $T>T_S$, $\langle \phi\rangle = 0$, signifying the $\beta$ phase, while for $T<T_S$ one stabilizes $\langle \phi \rangle  \neq 0$, obtaining a dimerized $\alpha$ phase\cite{Krumhansi_PRB_1989}. While for the $\beta$ phase, it is known that the system hosts quantum spin-Hall edge states (see Figure~\ref{Figure1}g), for the $\alpha$ phase, these gaps out with an effective dispersion as shown in Figure~\ref{Figure1}h. 
From our first principles calculations, the effective low-energy physics of the $\beta$ phase is given by $H \sim \hbar v_F k_y \sigma_x \otimes \tau_z $ where $\sigma$ represents the electronic spin, $\tau$ represents atomic orbitals, and $\hbar v_F$  is approximately $2.674{\rm{~eV.{\mathring{A}}}}$ (see Figure \ref{Figure1}g and Supplemental information for details). Given the $\alpha$ phase ($\langle \phi \rangle \neq 0$) gaps out these edge states, to obtain the low-energy physics, we couple the $\phi$ field to low-energy quantum spin Hall edge states, such that the complete Hamiltonian is 
\begin{equation}
    H = \hbar k_y \sigma_x \otimes \tau_z +  \lambda \phi \Big[ {\mathbb{1}} \otimes \tau_z + {\mathbb{1}} \otimes \tau_x \Big] + {\mathcal{L}}
\end{equation}
Here $\sigma_{x}$, $\tau_{x}$ and $\tau_{z}$ are Pauli matrices and ${\mathbb{1}}$ is $2 \times 2$ identity matrix. For $\tilde{a}<0$ when the $\phi$ takes a finite value, the effective band structure of the fermions gap out with a dispersion as shown in Figure~\ref {Figure1}(h).  Interestingly, for $\phi \neq 0$, in the regime of $-2<\lambda \phi <0$, the non-trivial edge modes exist (see Figure~\ref{fig_flucchannel}a), leading to a non-zero integer-quantized spin winding number. The details of the calculations, along with a lattice model for the tight-binding system, are presented in the supplemental information. The first principles band structure estimates $|\lambda \phi| \sim 14.849 \times 10^{-3}~\rm{eV}$.

We next pose how the presence of Fermionic couplings affects the displacing transition. Under a mean field self-consistent calculation, where the total free energy for both the scalar field and the Fermionic ground state energy is minimized\cite{Loon_sshspring_2021} (see Supporting information for details) we find that as a function of temperature $\tilde{a} =(T - T_S)$ and effective Fermion-$\phi$ coupling parameter $\lambda$, the fermions effectively renormalize the $\phi$ field by a linear term. This modifies the critical point $\tilde{a}$ from $zero$, for a wide range of parameters as seen in Figure~\ref{fig_flucchannel}b. Interestingly, the global minima $\phi_{\rm{min}}$ can correspond to a topological phase for a broad regime of the parameters $\tilde{a}$ and $\lambda$ when $-2<\lambda \phi_{\rm{min}}<0$ (see Figure~\ref{Figure 4}b). The effective band structure of the fermions is, in fact, topological, consistent with higher-order topology as reported in earlier literature\cite{npjQuantum_topologicalphasetransition_Bi4i4}. A spin-winding number calculation on the phase diagram also shows winding number $W_{\text{spin}}=1$ (see Supporting information).

The analysis above naturally leads to an explanation for the origin of low-frequency telegraphic noise. Near the phase transition between the $\alpha$ and $\beta$ phase, it is expected that domains of $\alpha$ phase coexist between those of $\beta$ phase, characteristic of a first-order transition (see schematic in Figure~\ref{fig_flucchannel}c). A typical topologically insulating region in the quasi-one-dimensional chain of length $l$, the difference in energies of the edge modes is given by $\Delta E^{\rm{edge}}= W \exp(-l/\xi)$
where $W$ is the spectrum bandwidth and $\xi$ is the localization length of the edge modes in the $\alpha$-domain. The localization length can be calculated from the band structure of the system via
\begin{equation}
    \xi= \frac{\hbar v_{F}}{\Delta E^{\rm{Surf}}_{\alpha}}
\end{equation}
where $v_{F}$ is the Fermi velocity of electrons near the $\beta$-domain (note that edge modes are at the boundary with $\beta$-domain) and $\Delta E^{\rm{surf}}_{\alpha}$ is the gap in surface states of the $\alpha$ phase. Interestingly, these edge modes act as trap centers for the conducting electrons and give rise to the generation-recombination process described by McWhorter\cite{Hooge_1981}.
So, the difference in energies of edge modes gives rise to two-level fluctuations in the system, resulting in the Lorentzian component of noise with a characteristic frequency $f_{c}$. Using First principle band structure (see Figure \ref{Figure1}g,h and Supplemental information for details) we obtain bandwidth to be $W \approx 0.22 {\rm{~eV}}$, Fermi velocity $\hbar v_{F} \approx 2.674 {\rm{~eV.{\mathring{A}}}}$, and $\Delta E^{\rm{Surf}}_{\alpha} \approx 0.025 {\rm{~eV}}$. Thus, matching $hf_c \sim
\Delta E_{\rm{edge}}$, a typical domain of size $l \sim 380~ {\rm{nm}}$ leads to $f_{c} \sim 20~{\rm{mHz}}$ which is consistent with experimentally observed values.

\subsection{Conclusion:}
We have provided signatures of topologically non-trivial surface states in the $\beta$ phase of Bi$_4$I$_4$ through direct local current imaging. Additionally, we have also demonstrated features of edge modes in the $\alpha$ phase of Bi$_4$I$_4$  as captured within the time domain resistance fluctuations. Moreover, these signatures are robust to hysteretic thermal cycles across the resistive transition. Our theoretical framework, combining first-principles study with an effective theory of phonons coupled to itinerant electrons, captures this displacive topological transition. Furthermore, we identify that the electron tunneling process across the edge modes about the typical domain sizes can capture the effective scattering time scale, which dominates the telegraphic noise. We report transport signatures of the topological states in both phases of Bi$_4$I$_4$ across the finite-temperature first-order phase transition. 

\section*{Acknowledgments}
D.S.R acknowledges DST-Inspire for providing fellowship and also acknowledges IACS, Kolkata for giving research opportunities. D.S.R thanks Sourav Sarkar for his technical help. S.K. acknowledges IACS, Kolkata, for PhD fellowship support. S.P. acknowledges support from IIT Kanpur Institute Fellowship and S.M. acknowledges support from PMRF Fellowship. T.D., S.D acknowledge UGC, India for fellowship. AA acknowledges funding from IITK Initiation Grant (IITK/PHY/2022010). Numerical calculations were performed on the workstations {\it Syahi} and {\it Wigner} at IITK. This work is supported by (i) CSIR. Grant Number: 03/ 1511/23/EMR-II. (ii) Department of Science and Technology, Government of India (CRG/2023/001100). S.K.P. thanks NFSG grant from BITS-Pilani, Dubai campus, which supported this research, and also gratefully acknowledges discussions about use of  WannierTools with Prof. Jincheng Zhuang and Zhijian Shi. 


\section*{Author contributions}
D.S.R, S.J. have grown the samples and performed all basic characterizations with inputs from S.K. The c-AFM measurements were performed by D.S.R, S.K., S.N. All the transport, noise measurements, and formal analysis were performed by D.S.R., S.K., and S.P, S.M, S.K.P, A.A. contributed to the development of the theory. S.K.P. performed the first-principles calculations. M.M. contributed to conceiving the idea and designing the experiment, data interpretation, and analysis. All the authors contributed to the data interpretation and writing the manuscript.

\section*{Competing financial interests}
The authors declare no competing interests.

\section*{Data and materials availability:}
The data presented in the manuscript are available from the corresponding author upon request. 

\pagebreak
\section*{References}
\bibliography{ref}

\pagebreak

\clearpage

\begin{center}
\Large{\underline{Supplementary information for}}\\
\LARGE{Signatures of emergent surface states across a displacive topological phase transition in Bi$_4$I$_4$}
\end{center}

\author{Deep Singha Roy$^{1,\S}$, Sk Kalimuddin$^{1,\S}$, Subrata Pachhal$^{2,\$}$, Saikat Mondal$^{2,\$}$, Soham Das$^{1}$, Sukanya Jana$^{1}$, Arnab Bera$^{1}$,  Satyabrata Bera$^{1}$, Tuhin Debnath$^{1}$, Ankan Bag$^{1}$, Souvik Pramanik$^{1}$, Sudipta Chatterjee$^{1}$, Sanjib Naskar$^{1}$,Shishir Kumar Pandey$^{3,4*}$, Adhip Agarwala$^{2\dagger}$, Mintu Mondal$^{1\ddagger}$}

\begin{enumerate}
    \item[] $^1$ School of Physical Sciences, Indian Association for the Cultivation of Science, Jadavpur, Kolkata 700032, India 
    \item[] $^2$ Indian Institute of Technology Kanpur, Kalyanpur, Uttar Pradesh 208016, India 
    \item[] $^3$ Department of General Sciences, Birla Institute of Technology and Science, Pilani-Dubai Campus, Dubai International Academic City, Dubai 345055, UAE
    \item[] $^4$ Department of Physics, Birla Institute of Technology and Science, Pilani, Hyderabad Campus, Jawahar Nagar, Kapra Mandal, Medchal District, Telangana 500078, India 
    \item[] $^{\S}$  These authors contributed equally to this work
    \item[] $^{\$}$  These authors contributed equally to this work
    \item[] $^*$ e-mail: shishir.kr.pandey@gmail.com; $^\dagger$ e-mail: adhip@iitk.ac.in; $^\ddagger$ e-mail: sspmm4@iacs.res.in
\end{enumerate}

\noindent\textbf{In this Supplementary information, we present the details of Bi$_4$I$_4$ single crystal growth, their detailed characterization, conductive AFM measurement, extended c-AFM data in $\beta$-Bi$_4$I$_4$ phase, and the details of the low-frequency resistance noise spectroscopy measurements. Details of computational methods employed, as well as some key findings of these calculations, are also provided.}

\makeatletter

\renewcommand\l@section{\@dottedtocline{1}{0em}{5.3em}}
\renewcommand\l@subsection{\@dottedtocline{2}{2.3em}{4.2em}}
\makeatother

\clearpage
\tableofcontents

\renewcommand\thesection{Section S\arabic{section}}
\setcounter{section}{0}

\makeatletter
\setcounter{figure}{0}
\renewcommand{\thefigure}{\textbf{S}\arabic{figure}}
\makeatother

\addtocontents{toc}{\protect\setcounter{tocdepth}{2}} 

\clearpage

\section{Bi$_4$I$_4$ single crystal growth and characterization}
\begin{figure*}[ht!]
\centerline{\includegraphics[width=1.0\textwidth]{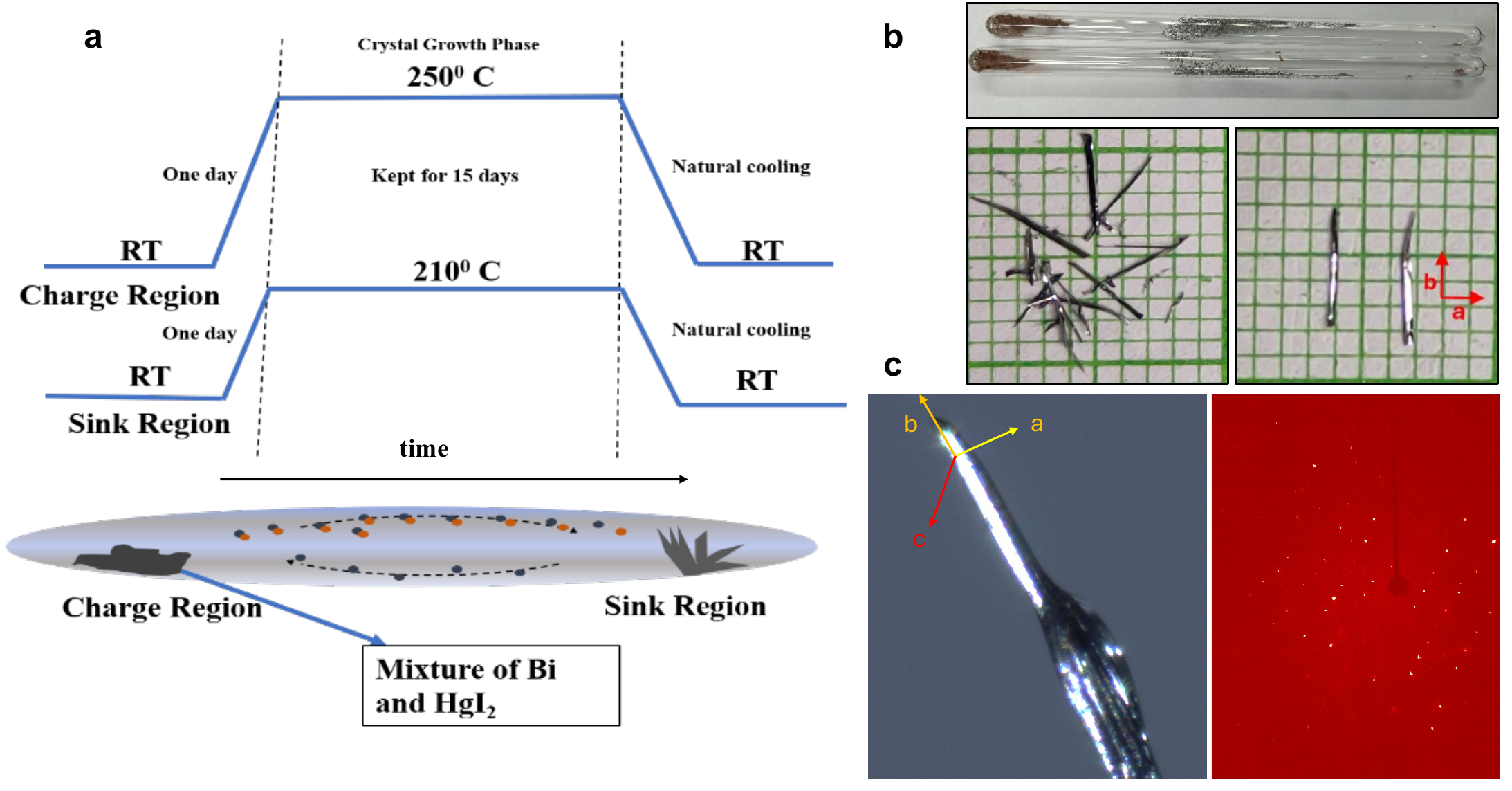}}
\caption{\textbf{Growth of Bi$_4$I$_4$ single crystals.} \textbf{a,} Schematic representation of the crystal growth process using the chemical vapor transport (CVT) method. Temperature profiles for the respective charge and sink regions are marked with time. \textbf{b,} Images of the quartz tubes with crystals inside (top) and as-grown crystals on a millimeter scale graph paper(bottom). \textbf{c,} Picture of the sample mounted on nylon sample holder (Left) and the corresponding Laue diffraction image.}
\label{Fig:Growth}
\end{figure*}

Bi$_4$I$_4$ single crystals are grown using the chemical vapor transport (CVT) technique. A mixture of Bi metal powder (purity 4N) and HgI$_2$ (purity 4N) in a 1:4 ratio of total mass 1.5~g is prepared and vacuum (~$10^{-4}$mbar) sealed in a quartz tube of length 25~cm. The evacuated ampule is then transferred to a two-zone furnace and kept for 15 days in a temperature gradient of 250$^{\circ}$C/ 210$^{\circ}$C\cite{autes2016novel_WTI}. Then it is cooled to room temperature (RT). The temperature profile with time is charted in Fig.~\ref{Fig:Growth}a. 
Needle-like single crystals with a typical size~5mm$\times$0.5mm$\times$0.1mm grow in the cold zone (see Fig.~\ref{Fig:Growth}b). Single crystal XRD is performed on the as-grown single crystal using a rotating crystal method with a Mo${K_\alpha}$ source at 270~K. The sample is mounted on a nylon loop (See~Fig.~\ref{Fig:Growth}c(Left)). Subsequently, a reflection image is depicted (Right). The $\alpha-$Bi$_4$I$_4$ is identified from the solved structure.

\section{Structural phases and Peierls-like distortion}
Here, we schematically present the intriguing details of the crystal structure of $\alpha$ and $\beta$ phases of Bi$_4$I$_4$, and their comparison with symmetry indicators. It is observed that the primary difference between the two phases is the in-plane inversion centers. There is no in-plane inversion center in the $\alpha$-phase. Whereas, the $\beta$ phase has an inversion center in the Bi metallic chain. This difference in the inversion symmetry induces a shift in the Bi$_4$I$_4$ stacking in the $\alpha$ phase that causes a displacive-like structural phase transition in the system. 
\begin{figure*}[ht!]
\centerline{\includegraphics[width=1.0\textwidth]{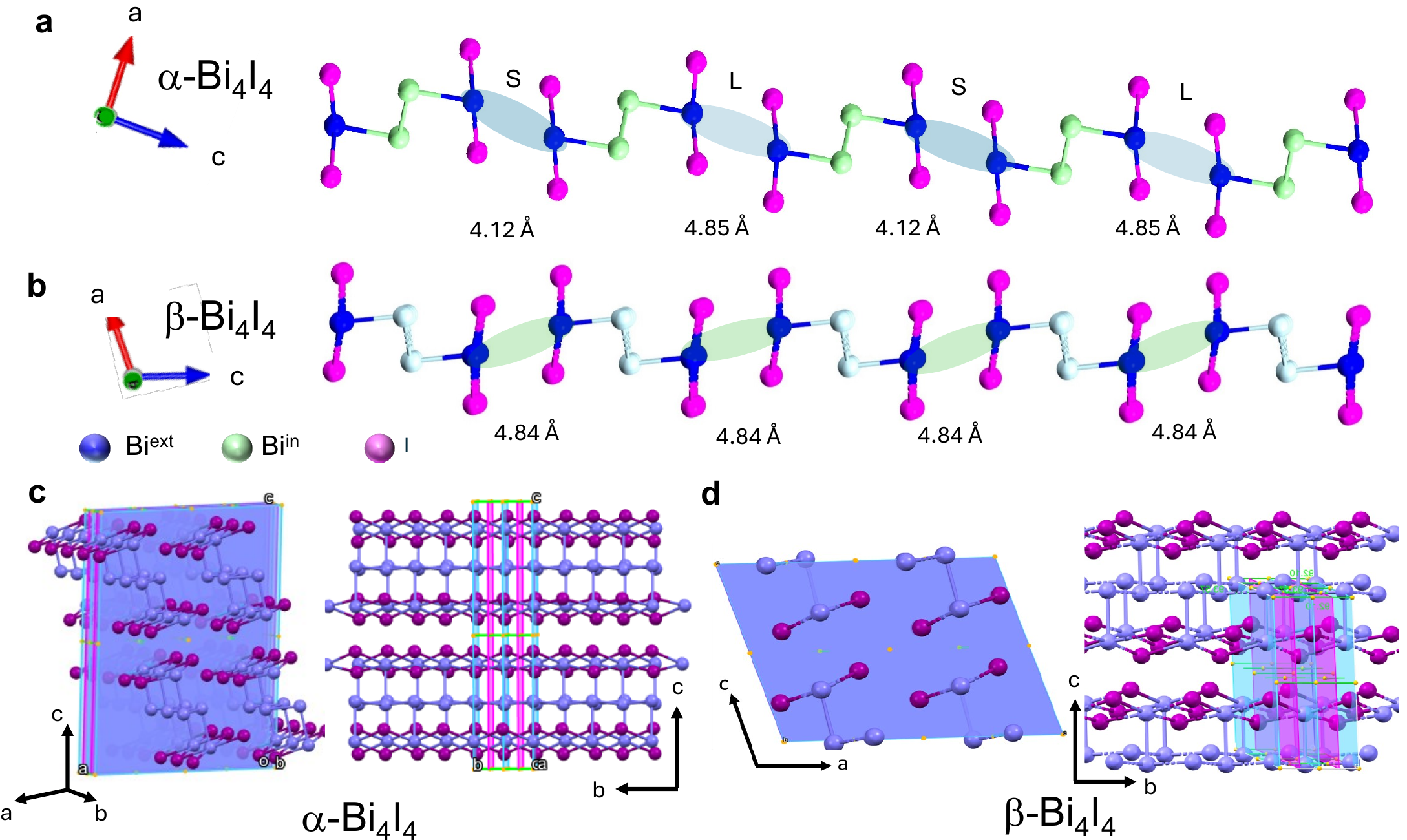}}
\caption{\textbf{Crystal structure, associated symmetries, and lattice dimerization of Bi$_4$I$_4$.} \textbf{a,b} Dimerization among the inter-layer peripheral Bi atoms in its $\alpha-$Bi$_4$I$_4$ phase. No distortion is observed for $\beta$-Bi$_4$I$_4$. The dimerization and periodicity along the stacking axis(\textbf{c}$^\prime$=\textbf{2c}-\textbf{a} direction) are marked by -L(long)-S(Short)-L(long)- sequence. \textbf{c}, Mirror symmetry and inter-layer inversion centers observed in $\alpha$-phase, but no in-layer inversion symmetry. \textbf{d,} For $\beta$ phase the in-layer inversion symmetry exists.} 
\label{Fig:Structure}
\end{figure*}
Now this subtle difference in crystal symmetry also induces a Peierls-like distortion dimerization along the van der Waals gap, i.e., the Bi-Bi dimerization occurs along \textbf{c}$^\prime$(i.e. \textbf{2c}-\textbf{a}) direction \cite{yoon_arxiv_2020}. This is understood by the identification of the distances between the Bi atoms residing at extreme positions in two consecutive Bi$_4$I$_4$ molecules. A periodic modulation of this separation, -Short-Long-Short-Long (i.e. -S-L-S-L-) occurs along the van der Waals gap. This periodic modulation is shown in Fig.~\ref{Fig:Structure}a,b. Those peripheral Bi atoms stretched along the van der Waals gap (i.e. \textbf{c}$^\prime$ direction) can be attributed as $Bi^{ext}_1$, and $Bi^{ext}_2$ respectively. $Bi^{ext}_1$-$Bi^{ext}_2$ dimerization creates a Peierls like scenario~Fig.~\ref{Fig:Structure}a. This creates a gap at the Dirac point on the (100), unraveling a stark difference in the band topology of the system\cite{2024_mass_acquisition_PRL}. Both phases are mirror symmetric. Mirror symmetry is on the (010) plane, cutting the Bi-Bi metallic chain perpendicularly ( See \ref{Fig:Structure}c,d ).

\section{Details of First-principles calculations}\label{sec_dft}
The electronic structure and topological band properties of bulk Bi$_4$I$_4$ are examined through Density Functional Theory (DFT) calculations using using projector-augmented wave method \cite{PAW, PAWpotentials1} as implemented within the Vienna $Ab$-$initio$ Simulation Package (VASP)~\cite{Kresse}. The internal coordinates of the primitive crystal structures of $\alpha$ and $\beta$-Bi$_4$I$_4$ are first optimized with energy and Hellmann-Feynman force convergence criteria of 10$^{-7}$~eV and 10$^{-3}$~eV/\AA{}, respectively. We have used the generalized gradient approximation (GGA)-based functionals, namely PBE~\cite{PBE} in our calculations with 5$d^{10}$6$s^2$6$p^3$ and 5$s^2$5$p^5$ valence configurations for Bi and I, respectively. Plane-wave cutoff energy 325~eV and 8 $\times$ 8 $\times$ 5 $\Gamma$-centered $k$-mesh for Brillouin zone sampling, Gaussian smearing width of 0.10~eV are considered in our calculations. In our calculations, the spin-orbit coupling (SOC) is considered at the self-consistent level.

\begin{figure*}[ht!]
\centerline{\includegraphics[width=1\textwidth]{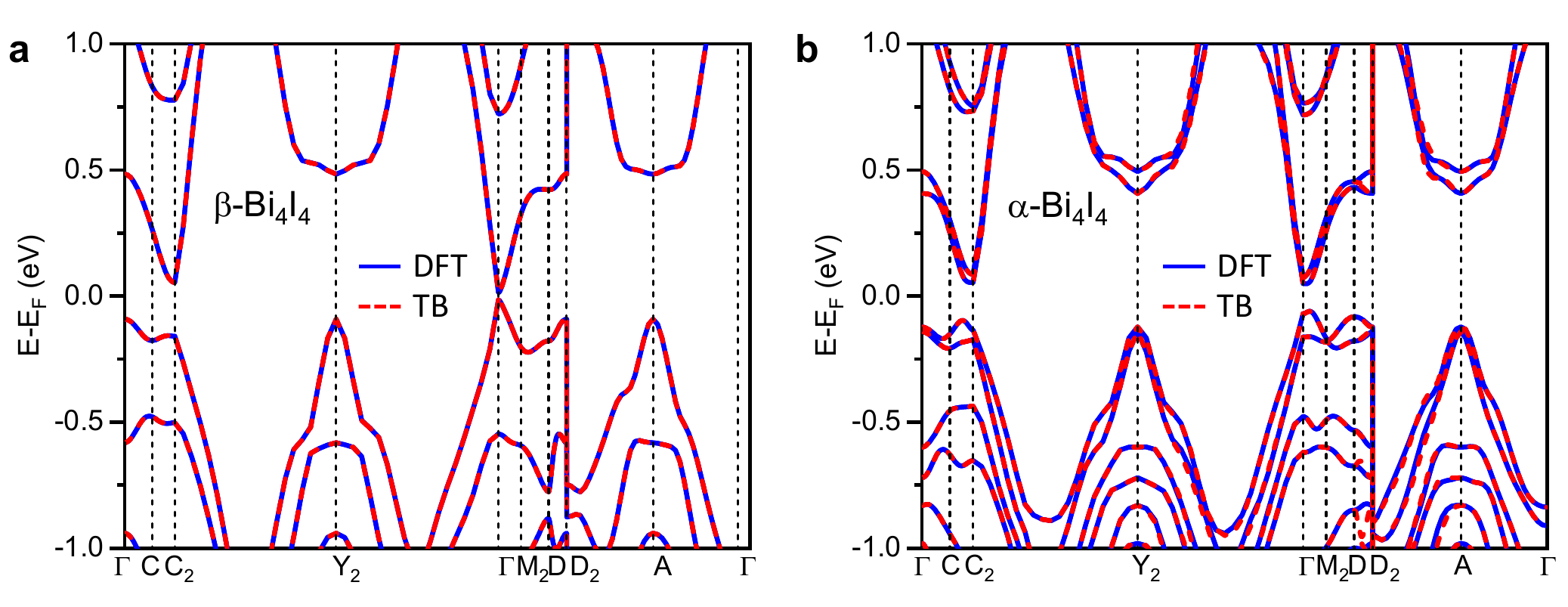}}
\caption{\textbf{Bulk band structure} for, \textbf{a} $\beta$-Bi$_4$I$_4$ and, \textbf{b} $\alpha$-Bi$_4$I$_4$ obtained with first-principles calculations.  Semi-metallic nature of $\beta$ phase is evident, while in $\alpha$ phase, a small bandgap $\sim$ 100~meV can be seen. TB parametrization using MLWFs is also shown in the respective plots. A good fit can be observed in both cases. Fermi energy is set to zero in all these plots.}
\label{Fig:Bulkband}
\end{figure*}

Once the structures were optimized, tight-binding (TB) parametrization of the DFT band structure was done using Maximally localized Wannier functions (MLWFs) as implemented in the WANNIER90 code v1.2~\cite{wannier90,Franchini2012}, considering $p$-orbitals of Bi and I atoms in the basis. Surface state calculations were then performed using the Green function approach as implemented in WannierTools V2.7.0~\cite{wt}. The strength of spin-orbit coupling is obtained by a two-step procedure. First, a non-magnetic Wannier function-based TB model, as well as SOC-included $ab$ $intio$ band structure, is obtained. In the second step, SOC strength is extracted from the fitting of this band structure to the TB model after explicitly adding the SOC term for Bi and I-$p$ orbitals into the model. This fitting is performed using the TBSOC package~\cite{tbsoc}. 

\subsection{Bulk band structures, surface states and estimation of electronic parameters}
In Fig.~\ref{Fig:Bulkband}a-b, we show the $ab$-$inito$ band structures of $\beta$ and $\alpha$-Bi$_4$I$_4$, respectively, obtained after including the SOC interaction. One can clearly observe a semi-metallic like behavior in the $\beta$ phase with a small band gap $\sim$ 60 meV while $\alpha$ has a bandgap of $\sim$ 130 meV. MLWFs-based TB parametrization of these band structures is also shown in the same figures for comparison. Bi and I $p$ orbitals are chosen to form the basis of the TB models. A good fit can be observed in both cases, which provides us with the confidence to reliably use these TB models for further calculations of surface states. We would like to note that the hopping interactions extracted from these TB models exhibit long-range behavior. 
\begin{figure*}[ht!]
\centering
\includegraphics[width=8 cm, height = 6 cm]{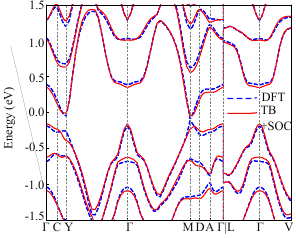}
\caption{{\bf SOC strength} estimation in $\beta$-Bi$_4$I$_4$ from fitting the SOC-included $ab$ $initio$ band structure (blue dashed lines)to the MLWFs-based TB model after explicitly adding the SOC term to the TB Hamiltonian (red solid lines) using TBSOC~\cite{tbsoc} package. A reasonably good fit can be observed. Fermi energy is set to zero.}
\label{Fig:soc_fit}
\end{figure*}

As explained above, the SOC strength was estimated in the $\beta$ phase from fitting the SOC-included $ab$ $initio$ band structure to the TB model after explicitly adding the SOC term to the TB Hamiltonian~\cite{tbsoc}. Bi and I -$p$ orbitals formed the basis of the TB model. One can find a reasonably good fit as shown in Fig.~\ref{Fig:soc_fit}. This provides an estimate of $\sim$ 1.35 eV. Since SOC is an atomistic property, we do not expect it to change in $\alpha$ phase.

\begin{figure*}[ht!]
\centerline{\includegraphics[width=1\textwidth]{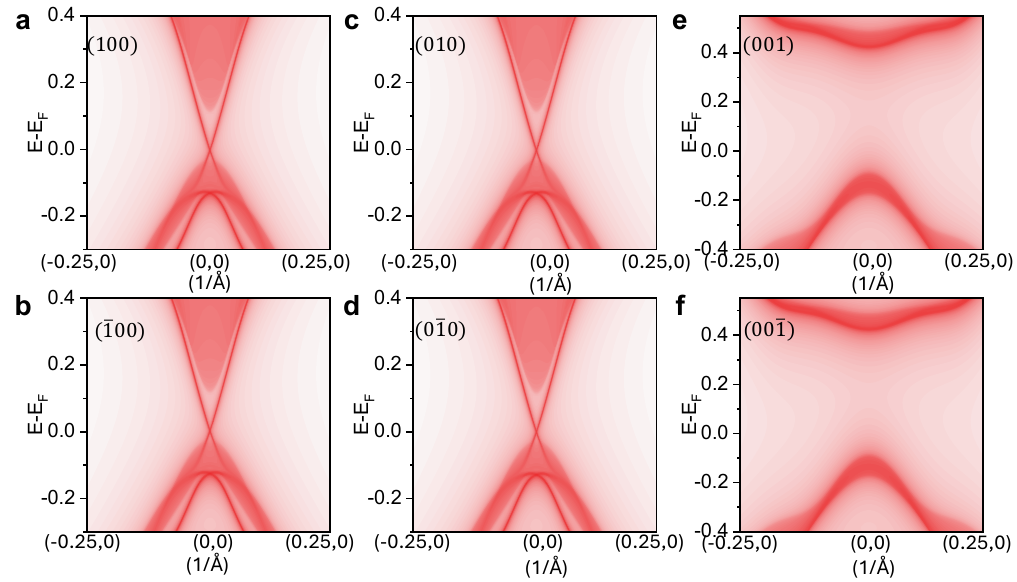}}
\caption{\textbf{Calculated surface states} of, \textbf{a, b,} (100) and ($\Bar{1}$00), \textbf{c,d} (010) and (0$\Bar{1}$0) and, \textbf{e,f} (001) and (00$\Bar{1}$) surfaces for $\beta$-Bi$_4$I$_4$ using Green's function method. The top surface (001) is gapped, while all the side surfaces are conducting.}
\label{Fig:SurfaceStates}
\end{figure*}

Using the TB models, we have performed the surface states calculations using Green's function approach as implemented in WannierTools~\cite{wt}. The (100) and ($\bar{2}$01) surface states of the $\beta$ and $\alpha$ -Bi$_4$I$_4$, respectively, are shown in Figure 1 of the main text. In addition to that, in Fig.~\ref{Fig:SurfaceStates} the surface state plots for all six surfaces, namely (100), ($\bar{1}$00), (010), (0$\bar{1}$0), (001) and  (00$\bar{1}$), are shown.  Consistent with the experimental finding, the top surface (001)/(00$\bar{1}$) is gapped, whereas all four of the side surfaces are conducting.

\section{c-AFM study of Bi$_4$I$_4$ flakes}
We use the c-AFM technique to acquire a nano-scale image of current distributions in freshly prepared crystalline flakes of Bi$_4$I$_4$. We exfoliate a pristine single crystal in a glove box in an Argon environment and transfer the flakes onto a p-Si wafer. Despite having an apparent quasi-1D character in Bi$_4$I$_4$ crystals, it typically cleaves along the 001 direction (i.e., ab plane)\cite{WTI_Composite_Wyel_zhang2016}. A higher exfoliation energy along the a-direction compared to the c-direction favors this exfoliation\cite{WTI_Composite_Wyel_zhang2016}. For the temperature-dependent measurement Si-wafer is placed on a copper holder cooled by liquid nitrogen. The temperature marked in the c-AFM scans of the main text is recorded using a Lakeshore 332 temperature controller. The scan is recorded using Ti/Ir coated silicon tip with a typical radius of 20~nm. For our experiment, we use a typical scan rate of 1~Hz with varying bias voltage, which is marked in each scan.

\begin{figure*}[ht!]
\centerline{\includegraphics[width=1\textwidth]{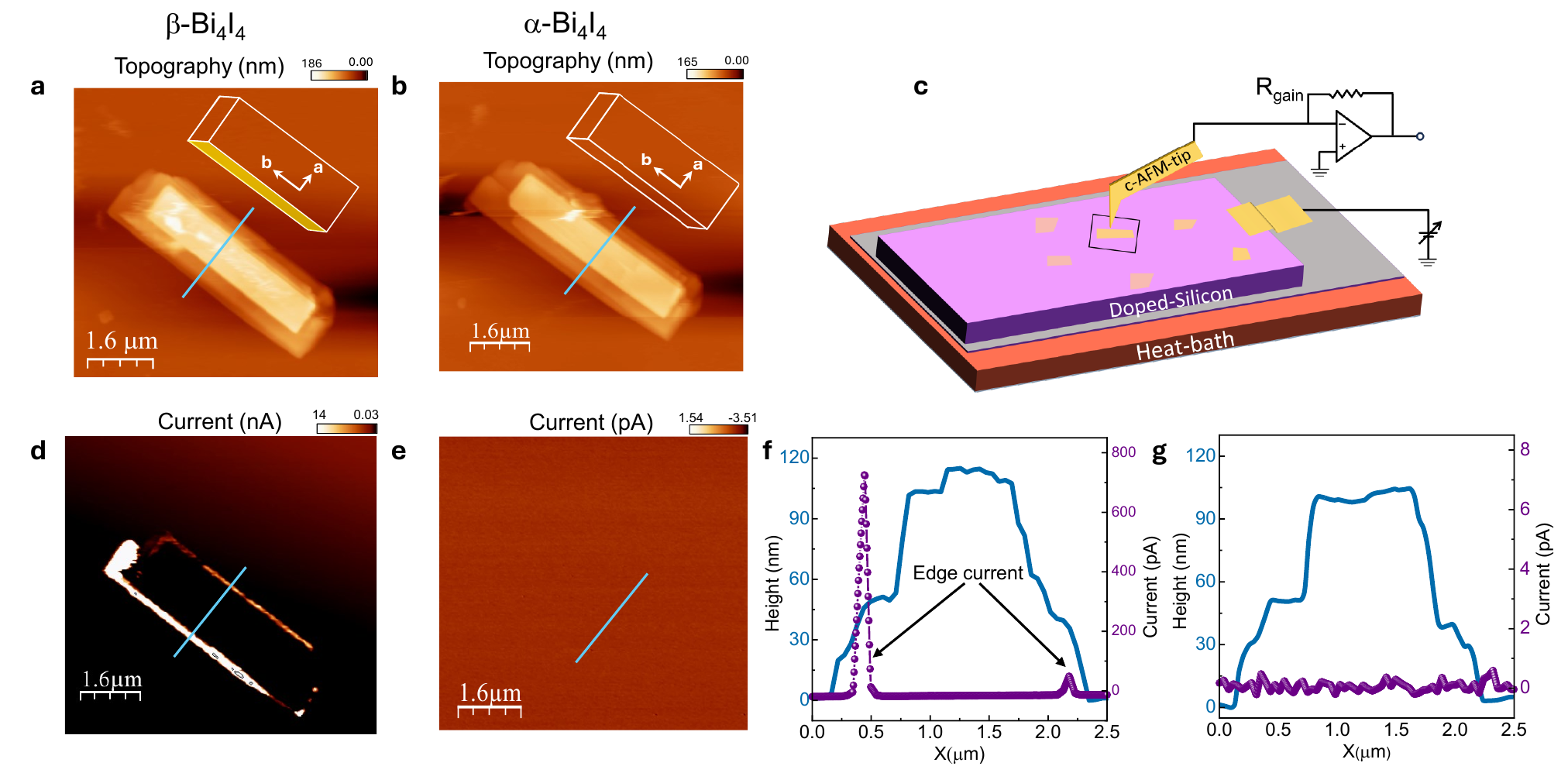}}
\caption{\textbf{Real space current mapping across the phase transition from $\beta$-Bi$_4$I$_4$ to $\alpha$-Bi$_4$I$_4$.} \textbf{a, b,} Topographic scan for $\beta$-Bi$_4$I$_4$ and $\alpha$-Bi$_4$I$_4$ (with a flake height $\sim$ 120nm) respectively. \textbf{c,} Schematic diagram of a cAFM measurement setup. \textbf{d,e} Current mapping of $\beta$ and $\alpha$ phases. \textbf{f,g} Respective plots of spatial variation of the current for both the phases.}
\label{Fig:CAFM1}
\end{figure*}

Our central observation using c-AFM is the vanishing edge current across the phase transition ($\beta$ to $\alpha$). Fig.~\ref{Fig:CAFM1}a,b shows the topographic scan of a Bi$_4$I$_4$ flake, and the inset shows a schematic representation of its orientation on p-Si wafer. Figure \ref{Fig:CAFM1}c is a schematic representation of the c-AFM measurement setup. Figure \ref{Fig:CAFM1}d,e shows the spatial distribution of current across the transition in two distinct phases. Also we have used bias voltage of 5V. Figure \ref{Fig:CAFM1}f,g shows the behavior of the height profile of the sample along with the corresponding current on a specific line-cross for two phases. This flake is different from the one presented in the main text. The surface states of the $\beta$ phase reappear even after multiple thermal cycling through the $\alpha$ phase, as captured through multiple samples. This clearly suggests the robustness of the system's topologically non-trivial nature across the room temperature phase transition.

\section{Robustness of surface conduction in spatial current of $\beta$-Bi$_4$I$_4$ flakes}
\begin{figure*}[ht!]
\centerline{\includegraphics[width=1.0\textwidth]{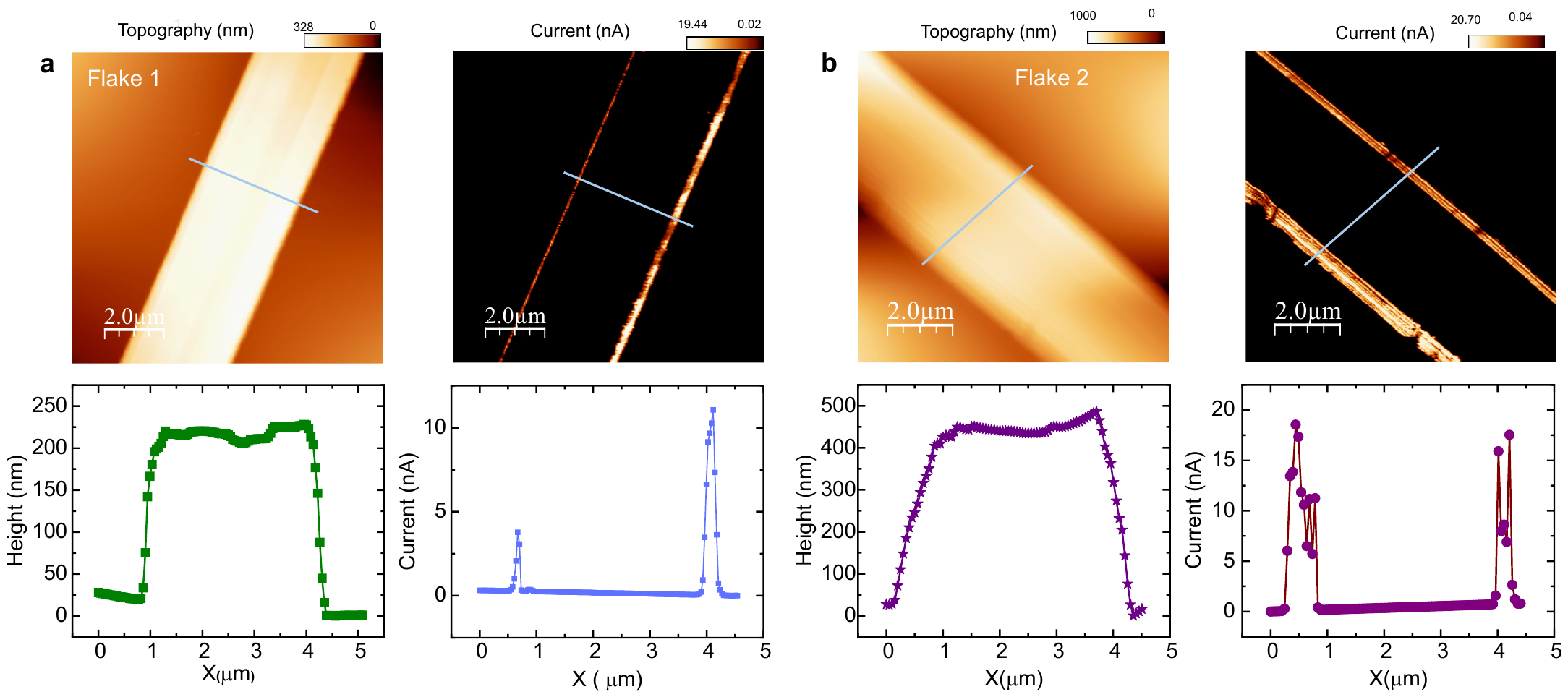}}
\caption{\textbf{Robust edge conduction in $\beta$-Bi$_4$I$_4$.} \textbf{a,} Extended data for topography (top left), spatial current map (top right), Height profile (bottom left), current variation (top left) along line cross for flake 1. \textbf{b-d} Represents the extended data for additional data three different flakes. The scan rate in both the flakes is 1~Hz, whereas the bias voltages are 4V and 5~V, respectively.}
\label{Fig:CAFM2}
\end{figure*}
The conducting surface states of $\beta-$Bi$_4$I$_4$ manifest in the form of a sharp contrast in current variation at side surfaces compared to the top surface (\textbf{a-b} plane). These edge conduction in the $\beta$ phase are reproducible as observed in multiple flakes from different crystals, as shown in Fig.~\ref{Fig:CAFM2}a-d. 

The presence of edge current is evident for all the flakes with more or less similar height at the room temperature.

\section{Lock-in based detection of resistance fluctuations}

\begin{figure*}[ht!]
\centerline{\includegraphics[width=1.0\textwidth]{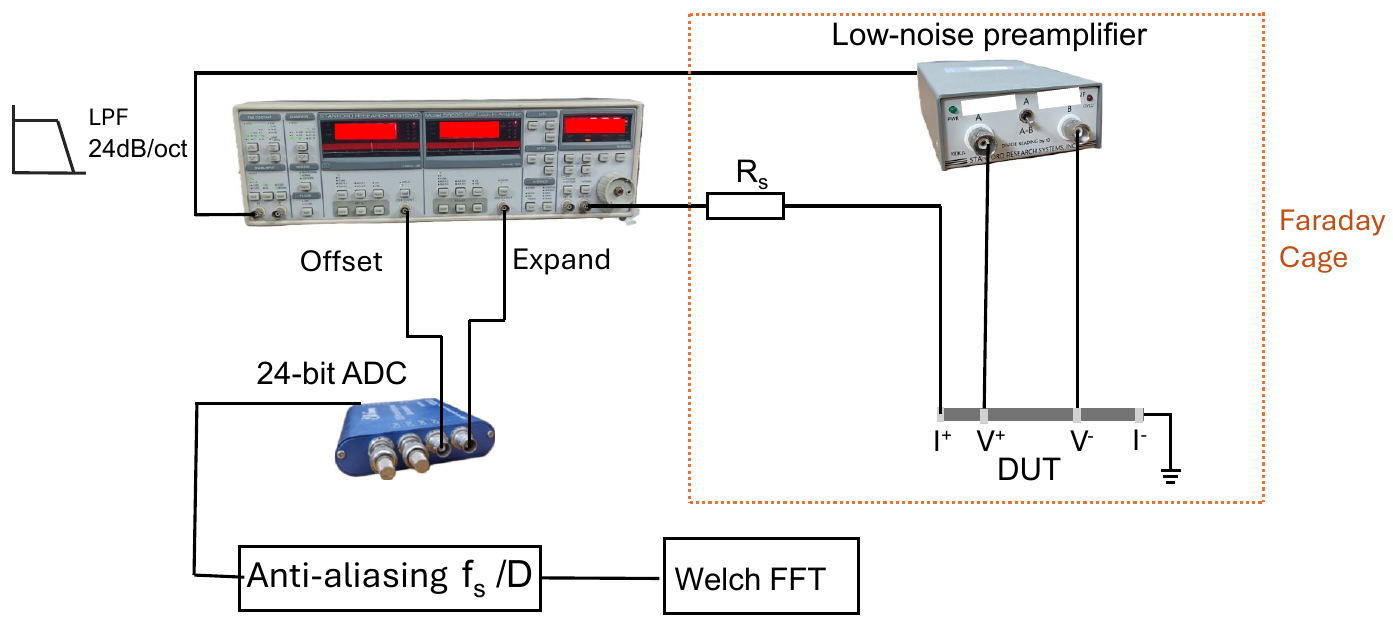}}
\caption{\textbf{Schematic representation 1/f Noise setup.} A typical four-probe noise setup with a lock-in playing the role of both source and sense. A 24-bit DAQ card for data acquisition, and the collected data is given a Welch FFT.}
\label{Fig:NoiseSchematic}
\end{figure*}

Here we have schematically represented a typical four-probe low-frequency noise spectroscopic setup in Fig. \ref{Fig:NoiseSchematic}. The experimental details are also available in ref. S. Kalimuddin et al PRL (2024)\cite{Kalimuddin_PRL}]. This lock-in-based measurement is carried out in interlock method using its own voltage source (sine out). We put a series resistance R$_s$ in current probe i.e. I$^+$. I$^-$ probe is connected to the common ground. We collect the voltage output from the middle probes (V$_A$ and V$_B$) in A-B mode. Afterwards, the signal is amplified using a voltage preamplifier (SR550). Then amplified output is fed into the lock-in A port. The output of the lock-in is collected using a 24-bit four-channel synchronous ADC, and the data are analyzed using an anti-aliasing filter. Data is stored by the ADC at a sampling rate of 1024 samples/sec for 4800 seconds. Finally, the timeseries data is given FFT to get the power spectral density (PSD). 

\section{Two level fluctuation}
\begin{figure*}[ht!]
\centerline{\includegraphics[width=0.7\textwidth]{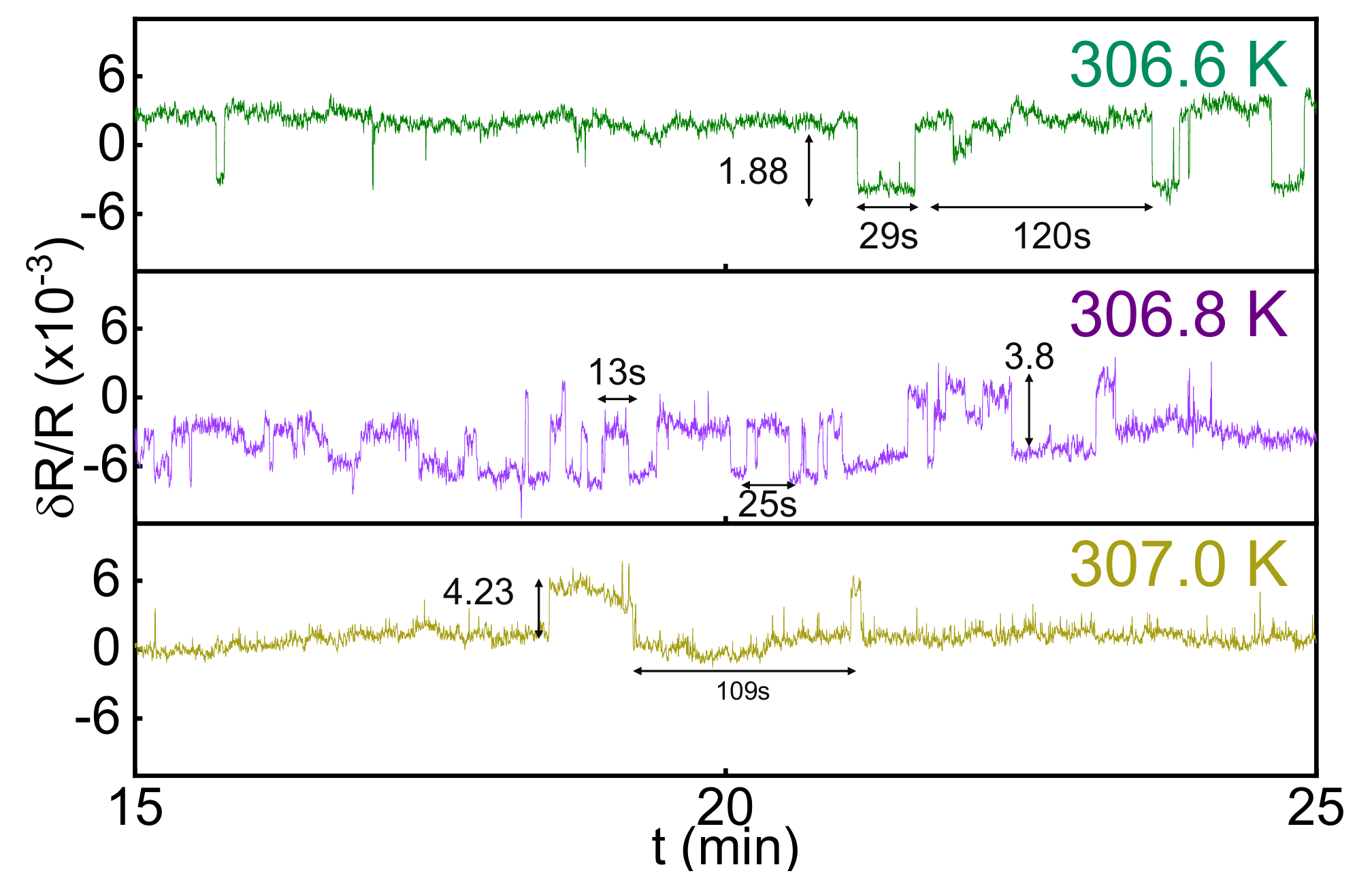}}
\caption{\textbf{Two level fluctuation.} Trimmed 10 minutes data representation of two-level fluctuation. }
\label{Fig:TwoLevel}
\end{figure*}

The two-level fluctuation in resistance time series (full data presented in the manuscript) is trimmed for a 10-minute scan and is shown in Fig.~\ref{Fig:TwoLevel}. For our case, we have observed a large fractional resistance step at the strong RTN region ($\delta$R/R($\%$)) varying from 0.1 to 0.4, which indicates a mesoscopic-like fluctuation in a bulk single crystal. This also signifies that the system has low disorder and is suggestive of transport dominated via shunted side surfaces.

\section{1/f and Lorentzian component superposition}
Prominent two-level fluctuations emerge near the phase transition($\alpha-\beta$ and vice-versa), evident in the time-traces of absolute ac-resistance. 
\begin{figure*}[ht!]
\centerline{\includegraphics[width=1.0\textwidth]{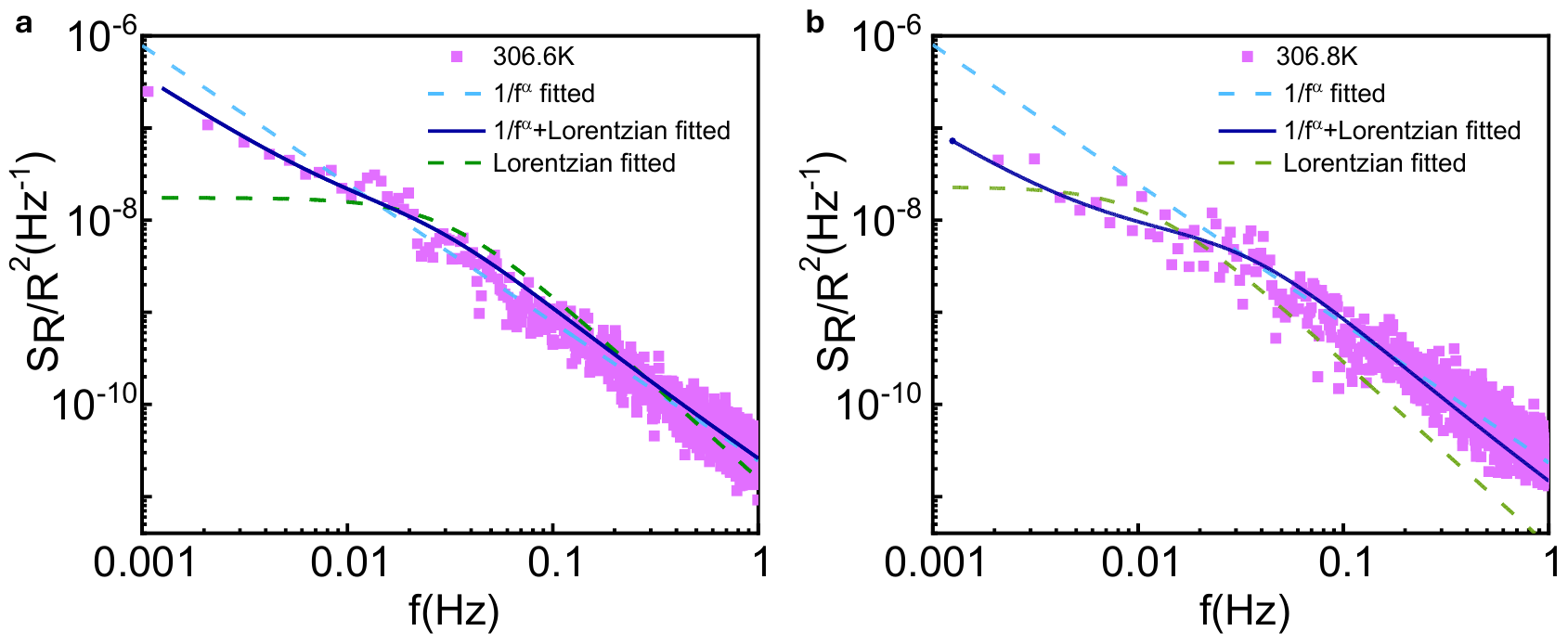}}
\caption{\textbf{Superposition of RTN Lorentzian and 1/f component.} \textbf{a,-b,} Fitting the power spectral density with 1/f$^\alpha$+Lorentzian and extracted the f$_c$ for temperatures 306.6K and 306.8K respectively.}
\label{Fig:Lorentzian}
\end{figure*} 
This, in turn, results in a strong Lorentzian feature superposed on 1/f-type spectra of resistance fluctuations. We therefore fit the PSD $A\left(\frac{1}{f^\alpha}\right)$+$B\left(\frac{f_c}{f^2+f_c^2}\right)$ and extracted the corner frequency of the Lorentzian associated with the two-level fluctuation. We fit the normalized PSD ($\frac{S_R}{R^2}$) as $\alpha$ and $\beta$ phases have different absolute resistance values. Now this $f_c$ parameter is a quantitative estimate of the timescale of the dynamics, and more importantly, it represents the energy barriers between the two distinct levels. Figure \ref{Fig:Lorentzian}a-b shows the PSD at two temperature values of 306.6~K and 306.8~K, respectively.

\begin{figure*}
\centering
\includegraphics[width=1.0\columnwidth]{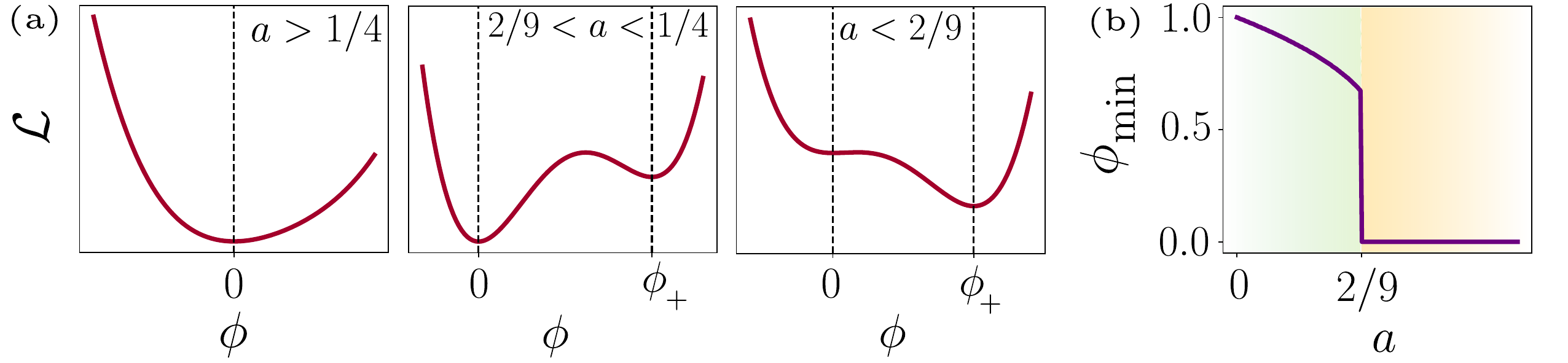}	
\centering
\caption{{\textbf{Free energy minima in classical $\phi^{3}$-theory:}} (a) Free energy density ${\mathcal{L}}$ (defined in Eq.~\eqref{eq_phi3}) as a function of $\phi$ for various values of $a$, where we find that there is a single minima for $a>1/4$, while there are two minima for $a<1/4$. When $2/9<a<1/4$, $\phi=0$ is the global minima and when $a<2/9$, $\phi=\phi_{+} \neq 0$ is the global minima. (b) Global minima $\phi_{\rm{min}}$ as a function of $a$, where $\phi_{\rm{min}} \neq 0$ only when $a<2/9$.}
\label{fig_phi_cube}
\end{figure*}

\section{Microscopic theory of displacive topological phase transition in Bi$_4$I$_4$}\label{sec_tpt}

\subsection{Revisiting displacive transition:}

Displacive or structural phase transitions are, in general, first-order in nature and can be captured via a scalar field $\phi$ of phononic origin. In the $\phi^{3}$-theory of first-order transition\cite{Krumhansi_PRB_1989}, the effective free energy density for the scalar field $\phi$ is given by
\begin{equation} \label{eq_phi3}
 {\mathcal {L}} = \frac{a}{2} {\phi^2}- \frac{1}{3} \phi^3 + \frac{1}{4} \phi^4,
\end{equation}
The minimization of $\mathcal {L}$ with respect to $\phi$ results in a single minima at $\phi=0$ when $a>\frac{1}{4}$ (see ~Fig.~\ref{fig_phi_cube}a). For $2/9<a<1/4$, although $\mathcal {L}$ has two minima, the global minima still remains at $\phi=0$ (see ~Fig.~\ref{fig_phi_cube}b). Whereas for $a<2/9$, $\phi$
at the global minima takes a finite non-zero value (see ~Fig.~\ref{fig_phi_cube}b), leading to a first-order transition. We replace $a=(\tilde{a} + 2/9)$ so that $\tilde{a}$ can be interpreted as the difference between the temperature $T$ and the critical temperature $T_{c}$, i.e. $\tilde{a} \propto (T-T_S)$. Here, the displacive transition occurs at the critical temperature $T=T_S$ (i.e. $\tilde{a}=0$). 

To explain the structural phase transition in Bi$_4$I$_4$, we consider the same $\phi^{3}$-theory where for $T>T_S$ \text{(i.e. $\tilde{a}>0$)}, $\langle \phi\rangle = 0$ represents the $\beta$ phase and for $T<T_S$ (i.e. $\tilde{a}<0$) the ordered phase with $\langle \phi \rangle \neq 0$, represents the $\alpha$ phase.

\begin{figure*}
\centering
\includegraphics[width=0.8\columnwidth]{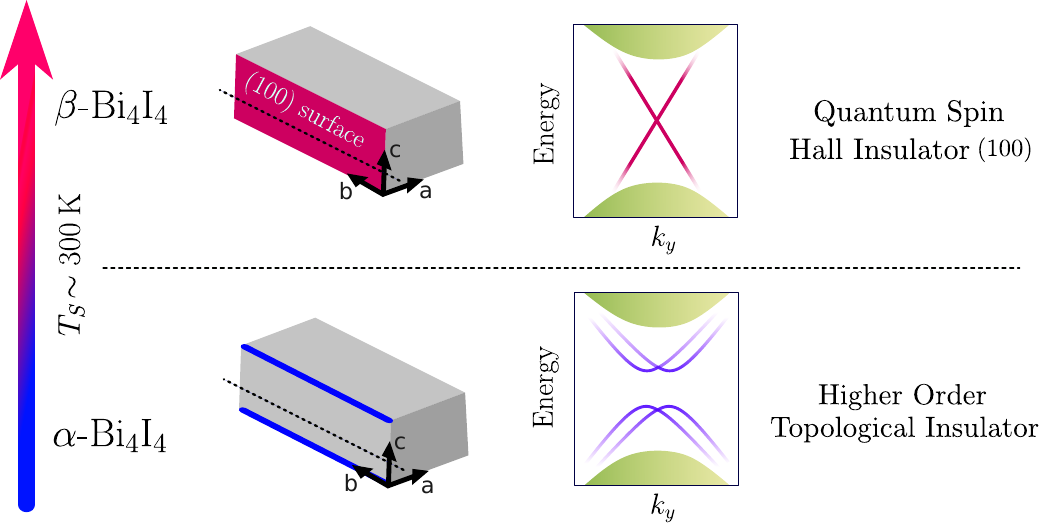}	
\centering
\caption{\textbf{Electronic phases of Bi$_4$I$_4$:} The (100) surface in $\beta$ phase hosts a linearly dispersing metallic edge state of the three-dimensional quantum spin Hall insulator, but becomes gapped in the $\alpha$ phase. The low-energy theory of the system can be reduced to one dimension along the dotted line.}
\label{fig_phases_sch}
\end{figure*}

\subsection{Electronic phase transition in Bi$_4$I$_4$:} Recent studies have shown that Bi$_4$I$_4$ is a quasi one-dimensional system with signatures of room temperature topological phase transition with $T_S \sim 300$K\cite{noguchi2019_WTI, PhysRevX.11.031042, npjQuantum_topologicalphasetransition_Bi4i4}. In ~Fig.~\ref{fig_phases_sch}, we illustrate both the low temperature and high temperature phases of Bi$_4$I$_4$ with their corresponding schematic low-energy dispersion predicted via first-principle studies\cite{2024_mass_acquisition_PRL,npjQuantum_topologicalphasetransition_Bi4i4}. While in the low temperature $\alpha$ phase, all the surfaces are gapped with hinge modes, leading to a higher-order topological insulator\cite{autes2016novel_WTI, PRX_3D_QSHI_Bi4I4yu2024, PhysRevX.11.031042, npjQuantum_topologicalphasetransition_Bi4i4, YuPRX2024, 2024_mass_acquisition_PRL, PhysRevX.11.031042, HOTI_noguchi2021evidence}, the high temperature $\beta$ phase hosts quantum spin Hall edge states in the (100) surface\cite{PRX_3D_QSHI_Bi4I4yu2024,PhysRevX.11.031042, noguchi2019_WTI}. In the low-energy sector, these gapless states are linearly dispersing along the \textbf{b} direction\cite{npjQuantum_topologicalphasetransition_Bi4i4} (which is also the chain direction of the quasi one-dimensional Bi$_4$I$_4$) and weakly coupled in the \textbf{c}-axis. All the other surfaces remain gapped\cite{PRX_3D_QSHI_Bi4I4yu2024, PhysRevX.11.031042}, which enables us to reduce the spatial dimension in the effective theory of the topological phase transition in this system. For instance, consider that the (100) surface is made of a bunch of one-dimensional systems (dotted line in ~Fig.~\ref{fig_phases_sch}) trivially coupled in the \textbf{c}-direction, then the effective theory reduces to one spatial dimension along the \textbf{b}-axis, where a metallic chain ($\beta$ phase) gets dimerized and give rise to a gapped phase ($\alpha$ phase) across the transition. 

In order to capture the low-energy physics of such a transition, we write the following electronic Hamiltonian coupled with the scalar field of the first-order transition, 
\begin{equation} \label{eq_Hel}
 H_{\text{el}} = \hbar v_F k_y \sigma_x \otimes \tau_z + \lambda \phi \big[ {\bOne} \otimes \tau_z + {\bOne} \otimes \tau_x \big],
\end{equation}
where $\sigma_{x}$, $\tau_{x}$ and $\tau_{z}$ are Pauli matrices and ${\bOne}$ is $2 \times 2$ identity matrix. From the first principle calculation (see~Fig.~\ref{sec_dft}), we estimate that the Fermi velocity $v_F$ is approximately given by $\hbar v_F = 2.674{\rm{~eV.{\mathring{A}}}}$. For now, we will consider $\lambda\phi$ to be a free parameter, but later we will show how $\phi$ can self-order from the full theory $H = H_{\text{el}} + \mathcal {L}$. While the first term of Eq.~Fig.~\ref{eq_Hel} gives linearly dispersing bands for $\phi=0$ mimicking the $\beta$ phase, the coupling parameter $\lambda$ opens up a gap when $\phi \neq 0$. To see if this gapped phase supports topology or not, let us now consider the following one-dimensional lattice Hamiltonian,
\beq
\mathcal{H}_{\text{el}}(k_y) = \sin (k_y) \sigma_x \otimes \tau_z + \big[\lambda\phi + 1 - \cos (k_y)\big]\bOne \otimes \tau_z + \big[ \lambda \phi + 1 - \cos (k_y)\big] \bOne \otimes \tau_x \label{eq_Kham}
\eeq
which reduces to the low-energy description as in Eq.~\eqref{eq_Hel} at the $\Gamma$ point ($k_{y}=0$). For simplicity, we ignore $\hbar v_F$ for now. Given $\lambda$ and $\phi$ are real, the symmetries of the Hamiltonian $\mathcal{H}_{\text{el}}(k_y)$ are, 
\begin{align}
	\text{Time-reversal symmetry},~ \mathcal{T} &: \sigma_y \otimes \bOne \mathcal{H}_{\text{el}}^{*}(k_y) \sigma_y \otimes \bOne = \mathcal{H}_{\text{el}}(-k_y);~\mathcal{T}^2=-1,\nonumber \\
	\text{Charge conjugation symmetry},~\mathcal{C}&: \sigma_y \otimes \tau_y \mathcal{H}_{\text{el}}^{*}(k_y) \sigma_y \otimes \tau_y = -\mathcal{H}_{\text{el}}(-k_y);~\mathcal{C}^2=1,\nonumber \\
	\text{Chiral symmetry}~\mathcal{S} &: \bOne \otimes \tau_y \mathcal{H}_{\text{el}}(k_y) \bOne \otimes \tau_y = -\mathcal{H}_{\text{el}}(k_y);~\mathcal{S}^2=1.\label{eq_symmetry}
\end{align}
\begin{figure*}
 \centering
\includegraphics[width=0.63\linewidth]{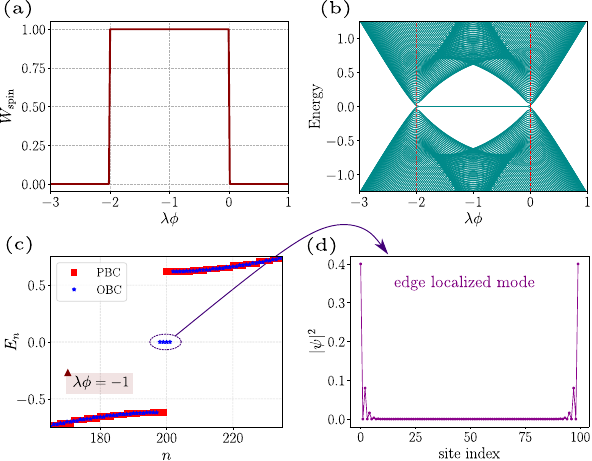}
 \caption{{\textbf{Topological character:}} (a) Winding number per spin texture (${W}_{\text{spin}}$) characterizes the topological phase of the Hamiltonian Eq.~\ref{eq_Kham} in the $\lambda \phi$ parameter space. (b) The energy spectrum of the Hamiltonian Eq.~\ref{eq_Rham} on an open chain of length $L=100$. (c) Energy eigenvalues of both the open and the periodic chain for $\lambda \phi=-1$. (d) Wave functions of the zero-energy states are localized at the two edges of the chain.}
 \label{fig_top}
\end{figure*}
Thus, the effective system belongs to the DIII symmetry class of tenfold classification, which exhibits topological properties in one dimension characterized by $\mathbb{Z}_2$ indices\cite{Altland_PRB_1997, Kitaevtenfold, chiu_classification_2016, Agarwala_AOP_2017}. However, because of spin-degenerate bands in the Brillouin zone, the calculation of topological indices such as winding number becomes tricky. Thus, we first break the degeneracy by adding a symmetry-respecting term $m\sigma_y\otimes\tau_x$ in the Hamiltonian in Eq.~\ref{eq_Kham}, with an infinitesimal value of $m$, and then calculate winding number per spin texture, ${W}_{\text{spin}}$. For all the bands in our system, ${W}_{\text{spin}}$ is quantized to unity in the parameter regime $-2< \lambda\phi<0$ as shown in ~Fig.~\ref{fig_top}(a), indicating a one-dimensional topological phase. To confirm the topological nature of the system, we rewrite the Hamiltonian given in Eq.~\ref{eq_Kham} in real space form, representing $\sigma = \{\uparrow, \downarrow\}$ as the actual spin and $\tau = \{A, B\}$ as an orbital label, 
\begin{align}
 H_{R} & = \sum_n \begin{pmatrix}
		c^{\dagger}_{n, \uparrow, A} & c^{\dagger}_{n, \uparrow, B} & c^{\dagger}_{n, \downarrow, A} & c^{\dagger}_{n, \downarrow, B}
	\end{pmatrix} \begin{pmatrix}
		1 + \lambda\phi & 1 + \lambda\phi & 0 & 0\\
		1 + \lambda\phi & -1 - \lambda\phi & 0 & 0\\
		0 & 0 & 1 + \lambda\phi & 1 + \lambda\phi\\
		0 & 0 & 1 + \lambda\phi & -1 - \lambda\phi
	\end{pmatrix} \begin{pmatrix}
	c_{n, \uparrow, A} \\ c_{n, \uparrow, B} \\ c_{n, \downarrow, A} \\ c_{n, \downarrow, B}
	\end{pmatrix} \notag \\
 & ~~~~~~~~~+ \sum_n \begin{pmatrix}
		c^{\dagger}_{n, \uparrow, A} & c^{\dagger}_{n, \uparrow, B} & c^{\dagger}_{n, \downarrow, A} & c^{\dagger}_{n, \downarrow, B}
	\end{pmatrix} \begin{pmatrix}
		-\frac{1}{2} & -\frac{1}{2} & -\frac{i}{2} & 0\\
		-\frac{1}{2} & +\frac{1}{2} & 0 & +\frac{i}{2}\\
		-\frac{i}{2} & 0 & -\frac{1}{2} & -\frac{1}{2}\\
		0 & +\frac{i}{2} & -\frac{1}{2} & +\frac{1}{2}
	\end{pmatrix} \begin{pmatrix}
	c_{n+\hat{x}, \uparrow, A} \\ c_{n+\hat{x}, \uparrow, B} \\ c_{n+\hat{x}, \downarrow, A} \\ c_{n+\hat{x}, \downarrow, B} 
	\end{pmatrix} + h.c.
\label{eq_Rham}
\end{align}
where $c^{\dagger}_{n, \uparrow/\downarrow, A/B}/c_{n, \uparrow/\downarrow, A/B}$ are the fermionic creation/annihilation operators for the corresponding spin and orbital label at $n_{th}$ site. It is evident from Eq.~\ref{eq_Rham} that the coupling of the phononic field $\phi$ to the itinerant fermions renders an onsite staggered potential and onsite hopping between A and B orbitals for both the spin species. Using numerical diagonalization, we find that the Hamiltonian in an open boundary condition hosts zero-energy modes for the parameter range $-2<\lambda \phi<0$, as is evident from the spectrum shown in ~\ref{fig_top}(b) and (c). Furthermore, we confirm that these zero-energy eigenstates are localized at the edges of the chain (see Fig.~\ref{fig_top}(d)).

So, in our model, as we tune $T$ below $T_S$ (i.e., $\tilde{a}<0$), $\phi$ takes a finite value and the gapless $\beta$ phase vanishes. However, for a finite $\phi$, the system hosts topological features mimicking the $\alpha$ phase only if $-2<\lambda \phi<0$. We will now show how the theory of the structural/displacive transition itself favours this specific range of the parameter $\lambda\phi$.

\begin{figure*}
 \centering
\includegraphics[width=0.85\linewidth]{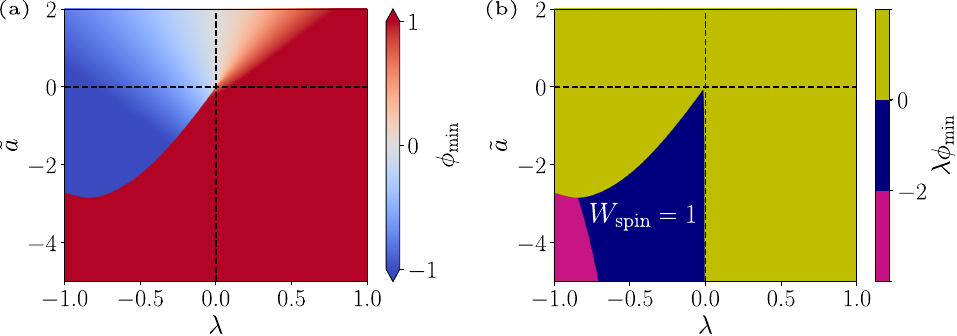}
 \caption{(a) The global minima $\phi_{\text{min}}$ of the total free energy density (with both electronic and classical part, see Eq.~\ref{eq_totalfreeenergy}) in the $\lambda$ - $\tilde{a}$ parameter space. (b) The region where $\lambda \phi_{\text{min}}$ is within the range $(-2, 0)$, electronic Hamiltonian become topological.}
 \label{fig_phimin}
\end{figure*}

\subsection{Self-ordering of the scalar field:} For the electronic Hamiltonian (see Eq.~\ref{eq_Kham}) at half filling (here two lowest bands will be filled), the ground state energy for a given momentum will be, 
\begin{align}
E(k_{y}, \phi, \lambda) = & -\sqrt{\big(\lambda \phi+1-\cos{k_y}\big)^{2}+\big(\lambda \phi +1-\cos{k_y}+\sin{k_y}\big)^{2}} \nonumber \\
&- \sqrt{\big(\lambda \phi +1-\cos{k_y}\big)^{2}+\big(\lambda \phi +1-\cos{k_{y}}-\sin{k_y}\big)^{2}}.
\end{align}
So for the full theory $H = H_{\text{el}} + \mathcal{L}$, total free-energy density is given by, 
\beq
F (\tilde{a}, \lambda, \phi) = \frac{1}{2\pi}\int_{-\pi}^{\pi} E(k_y, \phi, \lambda) dk_y + \frac{1}{2} \left(\tilde{a} + \frac{2}{9}\right)\phi^{2} -{{\frac{1}{3}\phi^3}} +\frac{1}{4} \phi^{4}.\label{eq_totalfreeenergy}
\eeq
Using numerical integration, we evaluate the total free energy density $F$ and minimize it with respect to $\phi$. A similar idea has been explored in ref\cite{Loon_sshspring_2021}. In the $\lambda$ - $\tilde{a}$ parameter space, the value of scalar field ($\phi = \phi_{\text{min}}$) at which $F$ has a global minimum is shown in Fig.~\ref{fig_phimin}(a). In Fig.~\ref{fig_phimin}(b), we illustrate the region in the parameter space where $-2< \lambda \phi_{\text{min}} < 0$; consequently, the electronic Hamiltonian is topological.

In our theory, the coupling parameter $\lambda$ has to be very small negative number, such that for $T>T_S~(\tilde{a}>0)$ the global minimum still appears at $\phi_{\text{min}}=0$ (see Fig.~\ref{fig_phimin}(a)) keeping the system in the gapless $\beta$ phase and at temperature below $T_S$ ($\tilde{a}<0$) it goes to a topological phase similar (see Fig.~\ref{fig_phimin}(b)) to the $\alpha$ phase. The gap at the $\Gamma$ point in the $\alpha$ phase calculated via First principle method (see Fig.~\ref{sec_dft}), give us $2\sqrt{2}|\lambda \phi| \approx 0.042$ eV, thus $|\lambda \phi| \sim 14.849 \times 10^{-3}~\rm{eV}$ which is reasonably small such that we can consider a very small negative $\lambda$ for our effective theory of the first order transition in Bi$_4$I$_4$.

\section{Resistivity fluctuation near the displacive transition}

As the Bi$_4$I$_4$ sample goes through a displacive topological transition from $\alpha$ phase to $\beta$ phase, the DC resistivity shows random telegraphic noise near the critical temperature. Below, we explain how this noise marks one of the direct evidence of edge mode formation in the system. 

\subsection{Edge modes as a source of resistivity fluctuation:} With the temperature approaching $T_S$, as a characteristic of the first-order phase transition, metallic domains of $\beta$ phase start to form on the gapped (100) surface of $\alpha$-Bi$_4$I$_4$ before eventually transforming to $\beta$-Bi$_4$I$_4$ at higher temperature. Such domain formation creates excessive edge modes at the boundary with the topological $\alpha$ phase. While inside the domains, the electrons move freely; at the boundary of the domains, the existence of edge modes works as randomly distributed trap centers. Similar to our microscopic theory of the system (see ~\ref{sec_tpt}), we reduce the domain landscape of the (100) surface into an effective one-dimensional picture such that, effectively, some randomly distributed metallic chunks appear on a topological insulator landscape (schematically shown in Figure 4c in the main text with its corresponding energetic landscape). In such an effective picture, the tunneling of electrons from one conducting domain to another via the edge modes near the Fermi energy ($E_F$) will give rise to electrical conductivity. However, due to the random variation in domain sizes, the number of edge modes or trap centers fluctuates, rendering noise in the conductivity or resistivity data. The mechanism of such random telegraphic noise was described by McWhorter using the generation-recombination process\cite{Hooge_1981, VANDAMME2005507}. Let us first revisit some key details of the McWhorter model in the following section.

\begin{figure*}
 \centering \includegraphics[width=0.45\linewidth]{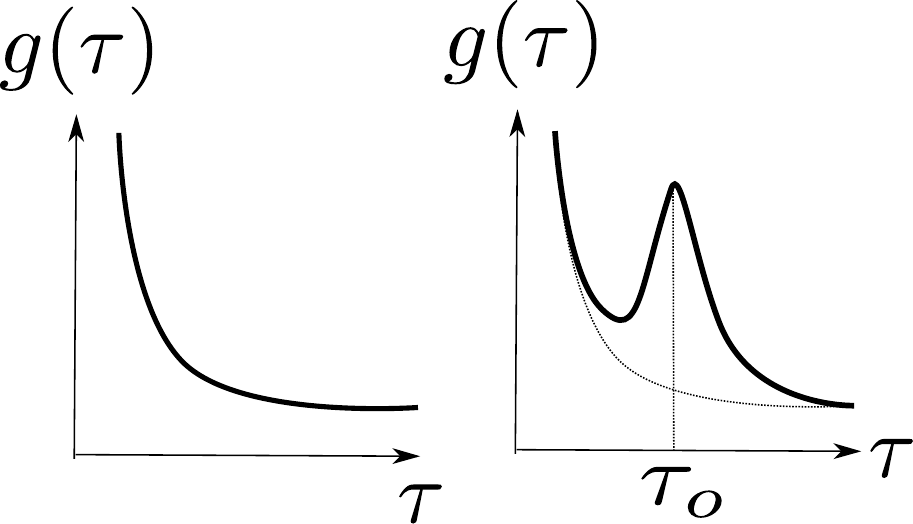}
 \caption{{\textbf{Weight factor for $1/f$ and Lorentzian noise:}} The weight $g(\tau)$ (see Eq.~\eqref{eq_gtau}) as a function of $\tau$ when $B=0$ and $B \neq 0$. Introduction of topological zero modes can lead to Lorentzian noise due to the intrinsic time scale $\tau_{0}$ associated with mid-gap states.}
\label{fig:scatteringrate}
\end{figure*}

\subsection{McWhorter model of noise:}
A time-dependent quantity (e.g. number of carriers) in the presence of noise is written as
\begin{equation}
 X(t)= \langle X \rangle + \Delta X(t),
\end{equation}
where $\Delta X(t)$ is the noise above the mean value $\langle X \rangle$. The autocorrelation function of the noise is then given by
\begin{equation}
 C(t) = \langle \Delta X(0) \Delta X(t) \rangle \sim \langle (\Delta X)^{2} \rangle \exp(-t/\tau),
\end{equation}
with $\tau$ being the timescale of the two-level fluctuations in the number of carriers, where the tunneling of carriers occurs randomly between the trap level and the conduction band. The Fourier transform of the autocorrelation function turns out to be
\begin{equation}
 S(f, \tau)= \int_{0}^{\infty} C(t) \cos(2\pi f t) dt \sim \langle (\Delta X)^{2} \rangle \frac{ \tau}{[1+ (2 \pi f \tau)^{2}]},
\end{equation}
leading to a Lorentzian spectrum of noise~\cite{Hooge_1981,VANDAMME2005507}. Now, if there are some traps at different distances $x$ from the surface, then the timescale $\tau$ can also be different for different distances $x$. Here, $\tau$ is assumed to grow exponentially with $x$ (where the probability of tunneling is inversely proportional to $\tau$), which implies $\tau \sim \exp(x/\lambda)$. Then the weight to the total $S(f)$ for the timescale $\tau$ is $g(\tau) = \frac{d n_{\rm{trap}}}{d \tau} = \frac{d n_{\rm{trap}}}{d x} \frac{d x}{d \tau} \sim 1/\tau$ where $n_{\rm{trap}}$ is the number of traps at any distance $x$. Thus, in this situation, it is found that
\begin{equation}
 S(f) = \int_{0}^{\infty} d \tau g (\tau) \frac{ \langle(\Delta X)^{2} \rangle \tau}{[1+ (2 \pi f \tau)^{2}]} \sim \frac{1}{f},
\end{equation}
leading to $1/f$ noise~\cite{Hooge_1981,VANDAMME2005507}. In the presence of both $1/f$ and Lorentzian components of noise, we assume the following form of the weight $g(\tau)$:
\begin{equation}\label{eq_gtau}
 g(\tau) =\frac{A}{\tau} + B \delta (\tau - \tau_{0}).
\end{equation}
The variation of $g(\tau)$ with $\tau$ when $B=0$ and $B \neq 0$ are shown in Fig.~\ref{fig:scatteringrate}. Here, $\tau_{0}$ is the time scale that corresponds to the fluctuation between the two levels. Thus, we obtain
\begin{align}
 S (f) &= \int_{0}^{\infty} d \tau g (\tau) S(f,\tau) \sim \frac{A}{f} + \frac{B \tau_{0}}{[1+ (2 \pi f \tau_{0})^{2}]},\nonumber \\
 \implies S(f) & \sim \frac{A}{f} +\frac{B^{\prime} f_{c}}{f^{2}+f_{c}^{2}}.
\end{align}
where $f_{c}=\frac{1}{2\pi \tau_{0}}$, $B^{\prime}=\frac{B}{2\pi}={\rm{constant}}$. Therefore, the presence of a Lorentzian component of noise having a characteristic $f_{c}$ indicates the existence of an intrinsic timescale $\tau_{0}$ that corresponds to two-level fluctuations in the system.

It is evident that around the transition from $\alpha$ phase to $\beta$ phase, resistivity contains both $1/f$ and Lorentzian components of noise (see main text). While the random distribution of edge modes gives rise to $1/f$ component, a typical topological region in the domain landscape (see Figure 4c in the main text) renders two-level fluctuation, causing Lorentzian components. Below, we estimate the characteristic $f_c$ of the Lorentzian component from the energy scale of the edge modes.

\subsection{Estimation of $f_{c}$:} Given a typical topologically insulating region of length $l$ (see Figure 4c in the main text), the difference in energies of the edge modes is given by 
\begin{equation}\label{eq_edge}
 \Delta E^{\rm{edge}}= W \exp(-l/\xi).
\end{equation}
where $W$ is the spectrum bandwidth for $\alpha$-${\rm{Bi}}_{4}{\rm{I}}_{4}$ and $\xi$ is the localization length of the edge modes. The localization length can be calculated from the band structure of the system via
\begin{equation}\label{eq_xi}
 \xi= \frac{\hbar v_{F}}{\Delta E^{\rm{Surf}}_{\alpha}},
\end{equation}
where $v_{F}$ is the Fermi velocity of electrons near the $\beta$-domain (note that edge modes are at the boundary with $\beta$-domain) and $\Delta E^{\rm{Surf}}_{\alpha}$ is the emergent surface gap of (1 0 0) surface in the $\alpha$ phase. As we described above, the difference in the energies of the edge modes gives rise to two-level fluctuations in the system, resulting in the Lorentzian component of noise with a characteristic frequency $f_{c}$. Using the relations given in Eq.~\ref{eq_edge}~and~\ref{eq_xi}, we estimate $f_{c}$ of the Lorentzian noise as follows:
\begin{align}
 h f_{c} & =\Delta E^{\rm{edge}}= W \exp(-l/\xi) \nonumber \\
\implies f_{c} &\approx \frac{W}{h}\exp\Bigg(-\frac{\Delta E^{\rm{Surf}}_{\alpha}l}{\hbar v_{F}}\Bigg).\label{eq_fc_domainsize}
\end{align}

\begin{figure*}
 \centering \includegraphics[width=0.55\linewidth]{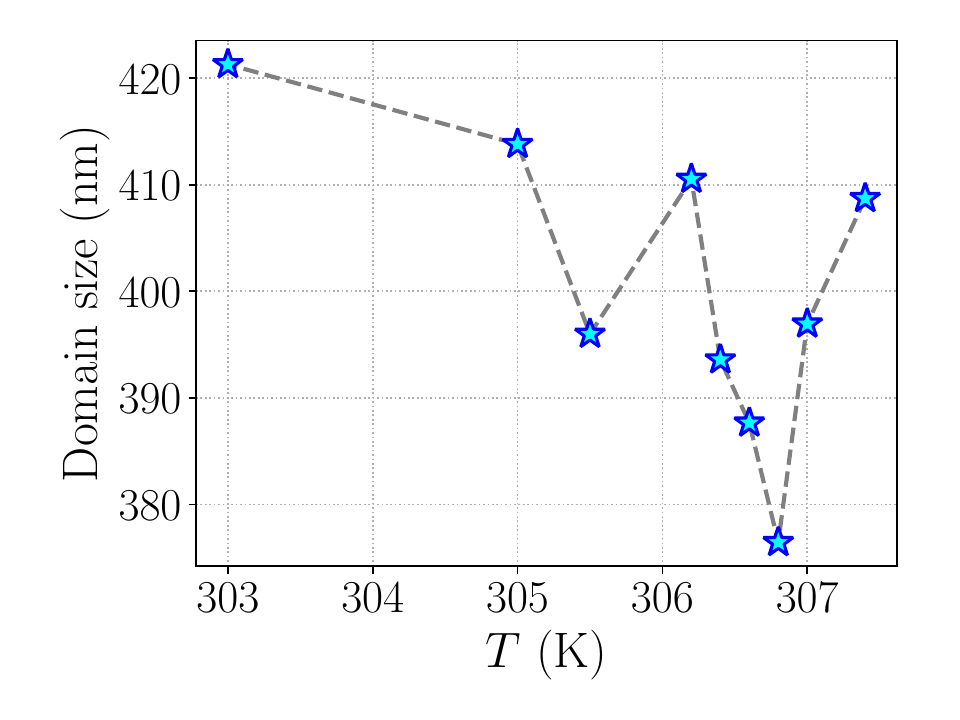}
 \caption{{\textbf{Domain sizes near displacive transition:}} Theoretical estimation of domain sizes of $\alpha$ phase from observed characteristic Lorentzian component $f_c$ of the two level fluctuation (see Eq.~\ref{eq_fc_domainsize}) near the critical temperature.}
\label{fig:Domainsize}
\end{figure*}
Using First principle band structure (see~\ref{sec_dft}) we obtain: the bandwidth of the low energy surface band to be $W \approx 0.22 {\rm{~eV}}$, the Fermi velocity to be given by $\hbar v_{F} \approx 2.674 {\rm{~eV.{\mathring{A}}}}$, and $\Delta E^{\rm{Surf}}_{\alpha} \approx 0.025 {\rm{~eV}}$. Given the experimentally observed $f_c$ values with temperature (see Figure 3e in the main text), from Eq.~\ref{eq_fc_domainsize}, we estimate the typical domain length scale $l$ of the $\alpha$-domain near the displacive topological phase transition. The calculated domain size variation is shown in Fig.~\ref{fig:Domainsize}. Thus a typical domain of size $l \sim 380~{\rm{nm}}$ leads to $f_{c} \sim 10~{\rm{mHz}}$. Although the calculation is performed considering {\it zero} temperature, since the bulk gap scale is higher than the transition temperature, the effective theory will remain valid.


\end{document}